\documentclass[12p]{iopart}

\usepackage{color}
\usepackage{graphicx}
\usepackage{iopams}
\usepackage{braket}
\usepackage[utf8]{inputenc}
\usepackage{url}
\usepackage{bbold}

\begin{document}

\title[Exponentially enhanced QC rate by multiplexing CV quantum teleportation]{Exponentially enhanced quantum communication rate by multiplexing continuous-variable teleportation}

\author{Andreas Christ$^1$, Cosmo Lupo$^2$, and Christine Silberhorn$^{1,3}$}
\address{\(^1\)Applied Physics, University of Paderborn, Warburger Straße 100, 33098 Paderborn, Germany}
\address{\(^2\)School of Science and Technology, University of Camerino,\\  via Madonna delle Carceri 9, I-62032 Camerino, Italy}
\address{\(^3\)Max Planck Institute for the Science of Light,\\  G\"unther-Scharowsky Straße 1/Building 24, 91058 Erlangen, Germany}

\ead{andreas.christ@uni-paderborn.de}

\date{\today}

\begin{abstract}
A major challenge of today's quantum communication systems lies in the transmission of quantum information with high rates over long distances in the presence of unavoidable losses. Thereby the achievable quantum communication rate is fundamentally limited by the amount of energy that can be transmitted per use of the channel. It is hence vital to develop quantum communication protocols that encode quantum information as energy efficiently as possible. To this aim we investigate continuous-variable quantum teleportation as a method of distributing quantum information. We explore the possibility to encode information on multiple optical modes and derive upper and lower bounds on the achievable quantum channel capacities. This analysis enables us to benchmark single-mode versus multi-mode entanglement resources. Our research reveals that multiplexing does not only feature an enhanced energy efficiency, leading to an \textit{exponential} increase in the achievable quantum communication rates in comparison to single-mode coding, but also yields an improved loss resilience. However, as reliable quantum information transfer is only achieved for entanglement values above a certain threshold, a careful optimization of the number of coding modes is needed to obtain the optimal quantum channel capacity.
\end{abstract}

\maketitle

\section{Introduction}
Quantum communication refers to the process of transferring quantum information between two parties commonly called Alice and Bob. This information transfer forms the cornerstone of many quantum  information technologies, most importantly quantum cryptography \cite{bennett_quantum_1984, ekert_quantum_1991}, enabling secure communication, quantum dense coding \cite{bennett_communication_1992}, boosting the  data rates with respect to classical transmission and quantum networking \cite{kimble_quantum_2008}. A major challenge in all these quantum communication protocols is to achieve high rates over long distances in the presence of unavoidable losses. For this purpose, we investigate continuous-variable (CV) quantum teleportation \cite{vaidman_teleportation_1994, braunstein_teleportation_1998}, as an established  method of transferring an unknown quantum state between two parties, using only entanglement and classical communication, which was originally introduced in 1993 by Bennett {\it et al.} \cite{bennett_teleporting_1993} in the discrete variable regime. 

In general, all quantum communication protocols are limited by the amount of energy that can be transferred between the sender (Alice) and the receiver (Bob) per use of the channel. Consequently, the challenge in quantum communication resides in encoding the information as energy efficiently as possible without sacrificing loss resilience. For this purpose, we expand the standard single-mode CV quantum teleportation protocol to incorporate multiplexing. Our research shows that by encoding the information on multiple instead of a single mode the information transfer is not only more energy efficient, leading to exponentially enhanced quantum channel capacity in comparison to the standard single-mode protocol, but it also features an enhanced loss resilience.

Furthermore, we propose a practical setup to implement the proposed multiplexing by encoding the information on ultrafast optical pulse modes\footnote{Ultrafast optical pulses are extremely short light pulses featuring durations in the femtosecond regime. Using these as carriers of quantum information enables the rapid succession of states in the transmission further boosting the quantum communication rate.}. There exists a wide variety of sources capable to create the required entangled states suitable for CV quantum teleportation, ranging from optical parametric oscillators \cite{laurat_experimental_2005, villar_generation_2005, laurat_entanglement_2005} over four-wave-mixing in optical fibers featuring a \(\chi^{(3)}\) nonlinearity \cite{loudon_squeezed_1987, levenson_generation_1985} to parametric down-conversion (PDC) in nonlinear \(\chi^{(2)}\) crystals \cite{rarity_observation_1987, wasilewski_pulsed_2006, lvovsky_decomposing_2007, wenger_pulsed_2004, anderson_pulsed_1997}. We employ --- without loss of generality --- an ultrafast pumped PDC source which creates a set of Einstein-Podolsky-Rosen  (EPR) states into ultrafast orthogonal frequency pulse modes, that can directly be applied for multiplexed quantum teleportation. 

We structured this paper into three main parts. In sections \ref{sec:single_mode_quantum_teleportation} and \ref{sec:single-mode_teleportation_analysis}, we review the standard single-mode CV quantum teleportation protocol to introduce all necessary concepts and formulas. Section \ref{sec:mm_eps_state_generation_and_teleportation} extends the standard protocol to include multiplexing. In section \ref{sec:quantum_channel_capacity_calculations}, we compare the achievable quantum communication rates in the multiplexed regime with the standard single-mode teleportation. Section \ref{sec:conclusion} concludes the paper and summarizes our findings.

\section{Single-mode CV quantum communication}\label{sec:single_mode_quantum_teleportation}
Before we present our multiplexed quantum communication protocol we first briefly review the established single-mode CV quantum teleportation scheme and the corresponding achievable quantum communication rates in order to introduce the required concepts and formulas.

\subsection{Teleportation as a quantum channel}

\begin{figure}[htb]
    \begin{center}
        \includegraphics[width=0.7\textwidth]{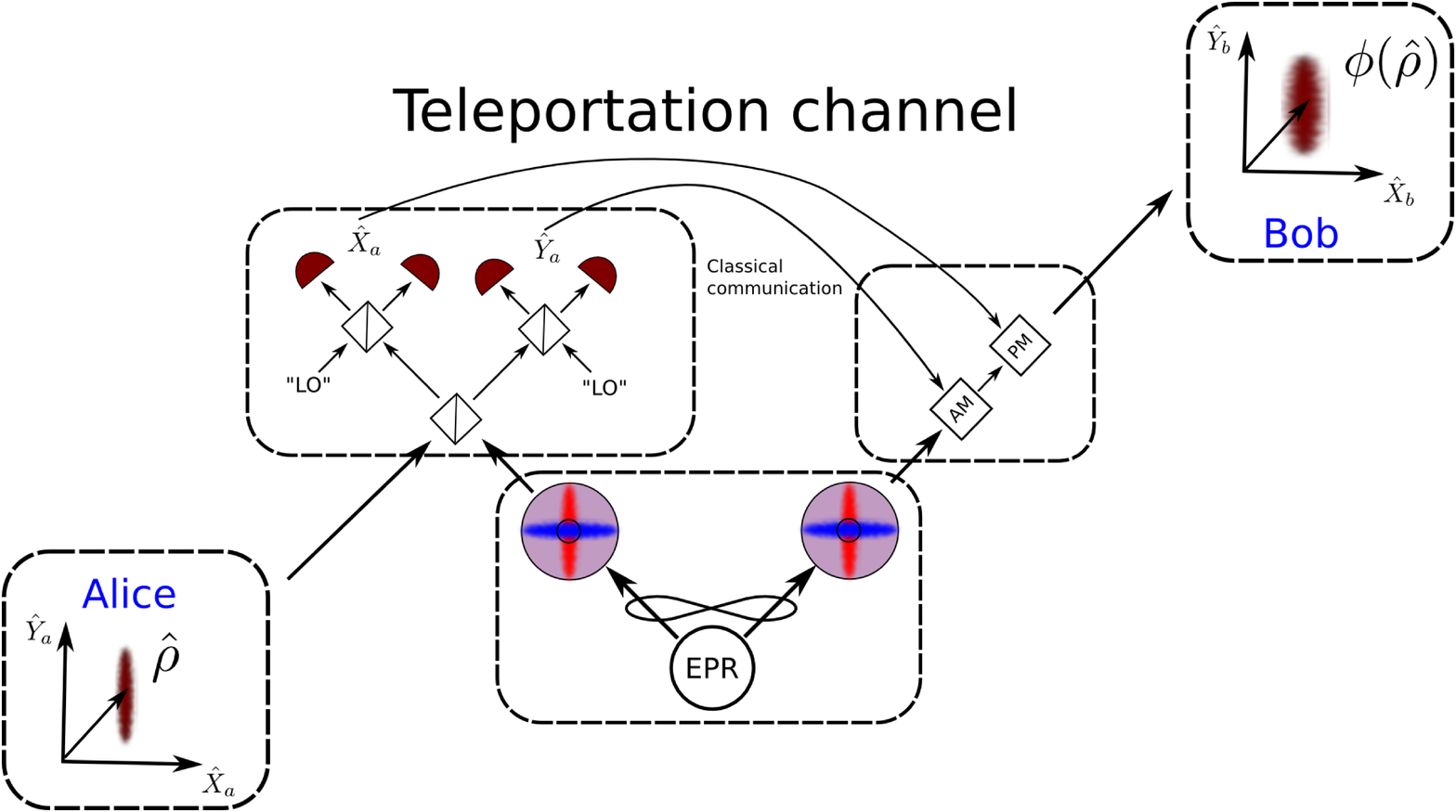}
    \end{center}
    \caption{Sketch of the standard single-mode CV teleportation protocol. An EPR state in conjunction with classical communication is used to transmit an unknown quantum state from Alice to Bob.}
    \label{fig:single-mode_teleportation}
\end{figure}

The standard single-mode CV quantum teleportation protocol \cite{vaidman_teleportation_1994, braunstein_teleportation_1998} is illustrated in figure \ref{fig:single-mode_teleportation}.
Alice intends to teleport a (unknown) quantum state \(\hat{\rho}\) from her side to Bob. To this aim, Alice and Bob share a bipartite entangled state --- in most cases a finitely squeezed EPR state --- associated with the operators \(\left\{\hat{a}, \hat{a}^\dagger\right\}\) on Alice's side, and \(\left\{\hat{b}, \hat{b}^\dagger\right\}\) on Bob's side obeying canonical commutation relations \(\left[\hat{a}, \hat{a}^\dagger\right] = \left[\hat{b}, \hat{b}^\dagger\right] = 1\). We denote the corresponding conjugate quadrature operators by \(\hat{q}_A = \left(\hat{a} + \hat{a}^\dagger\right) / \sqrt{2}\),  \(\hat{p}_A = \left(\hat{a} - \hat{a}^\dagger\right) / \imath \sqrt{2}\) and \(\hat{q}_B = \left(\hat{b} + \hat{b}^\dagger\right) / \sqrt{2}\),  \(\hat{p}_B = \left(\hat{b} - \hat{b}^\dagger\right) / \imath \sqrt{2}\) for Alice and Bob, respectively. 

The CV teleportation protocol works as follows: Alice first superimposes her part of the shared bipartite state --- we label it \(\hat{\chi}\) --- with the to be teleported state \(\hat{\rho}\). She then measures the resulting quantum system on her side and transmits the measurement result through classical communication to Bob. According to the data retrieved from Alice, Bob subsequently performs local operations on his part of the bipartite state \(\hat{\chi}\) and obtains the teleported state \(\hat{\rho}_{tel}\). 

In the scope of this paper, we are not interested in the details of the apparatus; hence we regard the whole protocol as a quantum channel which enables us to send a (unknown) quantum state \(\hat{\rho}\) from Alice to Bob. Then, we characterize the quantum channel defined by the teleportation protocol in terms of its quantum communication capacity. A reformulation of CV quantum teleportation as a quantum channel has been introduced by Ban {\it et al.} \cite{ban_continuous_2002}, extending that of Bowen and Bose \cite{bowen_teleportation_2001} on qubit teleportation. According to \cite{ban_continuous_2002} the CV teleportation protocol with arbitrary resources is formally described as a generalized thermalizing channel \(\phi(\hat{\rho}) = \hat{\rho}_{tel}\), in which thermal-like noise decreases the teleportation quality\footnote{For the qubit teleportation channel, the use of non ideal resources induces depolarization \cite{bowen_teleportation_2001}.}:
\begin{eqnarray}
    \phi(\hat\rho) = \int d x d y f(x,y) \hat{\mathcal{D}}(x,y) \hat{\rho} \hat{\mathcal{D}}^\dag(x,y) 
    \label{eq:thermalizing_quantum_channel_single_mode}
\end{eqnarray}
Here \(\hat{\mathcal{D}}(x,y)\) is the displacement operator
\begin{equation}
    \hat{\mathcal{D}}(x,y)(\hat{q} + \imath \hat{p})\hat{\mathcal{D}}^\dagger(x,y) = (\hat{q} -x) + \imath(\hat{p} - y),
\end{equation}
which shifts the input state \(\hat{\rho}\) in its quadratures \(\hat{q}\) and \(\hat{p}\) according to the function \(f(x,y)\) given by the structure of the channel. Consequently, Bob will receive the input state from Alice plus some extra phase-space displacements depending on the exact form of the CV teleportation. The input state from Alice is distorted from its original form. The exact structure of the mapping function \(f(x,y)\) is dependent on the shared bipartite state \(\hat{\chi}\) and is defined as
\begin{equation}
    \hspace{-1cm} f(x,y) = \mathrm{Tr} \left\{ \left[ \hat \mathbb{1} \otimes       \hat\mathcal{D}(x,y) \right] \left( |EPR^*\rangle\langle EPR^*| \right) \left[ \hat \mathbb{1} \otimes \hat\mathcal{D}^\dag(x,y) \right] \hat\chi \right\} \, ,
\label{eq:fxy_single_mode}
\end{equation}
where $|EPR^*\rangle$ denotes the not-normalized $EPR$ state
\begin{equation}
|EPR^*\rangle = (2\pi)^{-1/2}  \int_{-\infty}^\infty dq \, |q\rangle_A |q\rangle_B \, ,
\label{eq:not-normalized_EPR state}
\end{equation}
and $|q\rangle_A$, $|q\rangle_B$ are the eigenstates of the quadrature operators, $\hat q_{A} |q\rangle_A = q |q\rangle_A$, $\hat q_{B} |q\rangle_B = q |q\rangle_B$. 

Perfect teleportation is achieved for an infinitely squeezed  EPR state $\hat\chi = |EPR\rangle\langle EPR|$, which yields $f(x, y)=\delta(x)\delta(y)$. Hence, the input state \(\hat\rho\) from Alice is transmitted to Bob with unit fidelity, $\phi(\hat\rho)=\hat\rho$.

\subsection{CV teleportation with Gaussian resources}
In the remainder of this paper, we restrict ourselves to a Gaussian resource \(\hat{\chi}\) shared between Alice and Bob, as it is the case for the most common CV entangled state, the EPR state. The Gaussian state \(\hat{\chi}\) is conveniently described in the Wigner function representation:
\begin{eqnarray}
    \nonumber
    &W_\chi(q_A, p_A, q_B, p_B) = \frac{1}{\left(2 \pi\right) \sqrt{\mathrm{det} \gamma}} \exp\left[ -\frac{1}{2} (\xi - m) \gamma^{-1} (\xi - m)^T\right] \\
&= G_{(m,\gamma)}(q_A, p_A, q_B, p_B),
    \label{eq:gaussian_2_mode_wigner_function}
\end{eqnarray}
where \(\xi\) is defined as the vector \(\xi = (q_A, p_A, q_B, p_B )\), \(m\) labels the first-order moments and \(\gamma\) the second-order moments or covariance matrix (CM) of the state \(\hat\chi\), which completely characterize the Gaussian state. We have introduced the short-hand notation \(G_{(m,\gamma)}\) in \eref{eq:gaussian_2_mode_wigner_function}, where \(G\) marks the function as Gaussian in its variables, and the subscripts \(m\) and \(\gamma\) inside the brackets identify the first and second-order moments of the state.

The first step towards evaluating the output state of the teleportation channel is to derive the explicit form of the noise function $f(x, y)$ for a given Gaussian teleportation resource $\hat\chi$. Starting from the general form of \(f(x, y)\) in \eref{eq:fxy_single_mode} the function is given by the convolution integral
\begin{eqnarray}
    \fl \qquad f(x, y) = \pi \int d\xi \, W_{EPR^*}(q_A, p_A, q_B-x, p_B-y) \, G_{(m,\gamma)}(q_A,p_A, q_B,p_B),
    \label{eq:fxy_general_single_mode}
\end{eqnarray}
where $d\xi=d q_A \, d p_A \, d q_B \, d p_B$ and $W_{EPR^*}$ denotes the Wigner function of the not-normalized EPR state in \eref{eq:not-normalized_EPR state}.

To compute the convolution integral in \eref{eq:fxy_general_single_mode}, it is convenient to change to the collective quadratures $(q_-, p_-, q_+, p_+)$, defined as
\begin{eqnarray}
q_\pm  :=  \frac{q_A \pm q_B}{\sqrt{2}} \, , \qquad
p_\pm  :=  \frac{p_A \pm p_B}{\sqrt{2}} \, .
\end{eqnarray}
In terms of the collective variables, the Wigner function of the teleportation resource \(\hat{\chi}\) now reads
\begin{equation}
\tilde W_\chi(q_-, p_-, q_+, p_+) = G_{(\tilde m,\tilde\gamma)}(q_-, p_-, q_+, p_+) \, ,
\end{equation}
where $\tilde m = (m_{q_-}, m_{p_-}, m_{q_+},m_{p_+}) = m R$, with
\begin{eqnarray}
R = \frac{1}{\sqrt{2}} \left(\begin{array}{cc}
\mathbb{1}_{2}  & \mathbb{1}_{2} \\
-\mathbb{1}_{2} & \mathbb{1}_{2}
\end{array}\right) \, , 
\end{eqnarray}
$\mathbb{1}_{2}$ being the unit matrix of size $2$, and
\begin{eqnarray}
\tilde \gamma = R^t \gamma R = \left( \begin{array}{cccc}
\tilde\gamma_{q_-q_-} & \tilde\gamma_{q_-p_-} & \tilde\gamma_{q_-q_+} & \tilde\gamma_{q_-p_+} \\
\tilde\gamma_{p_-q_-} & \tilde\gamma_{p_-p_-} & \tilde\gamma_{p_-q_+} & \tilde\gamma_{p_-p_+} \\
\tilde\gamma_{q_+q_-} & \tilde\gamma_{q_+p_-} & \tilde\gamma_{q_+q_+} & \tilde\gamma_{q_+p_+} \\
\tilde\gamma_{p_+q_-} & \tilde\gamma_{p_+p_-} & \tilde\gamma_{p_+q_+} & \tilde\gamma_{p_+p_+} \\
\end{array}\right) \, .
\end{eqnarray}
In terms of the collective variables, the Wigner function of the not-normalized EPR state in \eref{eq:not-normalized_EPR state} reads
\begin{equation}
\tilde W_{EPR^*}(q_-,p_-,q_+,p_+) = 2\pi \, \delta(q_-) \delta(p_+) \, .
\end{equation}
We arrive at the final form of the mapping function \(f(x,y)\) for shared Gaussian resources
\begin{eqnarray}
    \nonumber
    f(x,y) & = & \frac{1}{2} \int d\xi \, \delta(q_{-}+x/\sqrt{2}) \delta(p_{+}-y/\sqrt{2}) G_{(\tilde m,\tilde\gamma)}(q_-, p_-, q_+, p_+) \\
    \nonumber
    & = & \frac{1}{2} G_{(m_f,\gamma_f)}(x/\sqrt{2}, y/\sqrt{2}) \\
    & = & G_{(\sqrt{2}m_f,2\gamma_f)}(x,y) \, ,
\end{eqnarray}
where $m_f = (\tilde m_{q_-},\tilde m_{p_+})$ and
\begin{eqnarray}\label{n-CM}
    \gamma_f = \left(\begin{array}{cc}
    \tilde\gamma_{q_- q_-} & \tilde\gamma_{q_- p_+} \\ 
    \tilde\gamma_{p_+ q_-} & \tilde\gamma_{p_+ p_+}
    \end{array}\right) \, .
\end{eqnarray}
This gives us a convenient closed formula for \(f(x,y)\) defined by the first moments \(m\) and CM \(\gamma\) of the shared resource \(\hat\chi\) between Alice and Bob. In particular, given a Gaussian state $\hat\rho$ on Alice's side with Wigner function 
\begin{equation}
W_{\rho}(q,p) = G_{(m_\rho,\gamma_\rho)}(q,p) \, ,
\end{equation}
the teleported state $\phi(\hat\rho)$ arriving at Bob's side evaluates to 
\begin{eqnarray}
    \nonumber
    W_{\phi(\rho)}(q,p) & = & \int d x\, d y \, f(x, y)\, G_{(m_\rho,\gamma_\rho)}(q - x, p - y) \\
    \nonumber
    & = & \int d x\, d y\, G_{(\sqrt{2}m_f,2\gamma_f)}(x, y) G_{(m_\rho,\gamma_\rho)}(q-x,p-y) \\
    & = & G_{(m_\rho+\sqrt{2}m_f,\gamma_\rho+2\gamma_f)}(q, p) \, .
    \label{eq:teleported_state_gaussian_framework_single_mode}
\end{eqnarray}
Equation \eref{eq:teleported_state_gaussian_framework_single_mode} fully determines the CV teleportation process in the Gaussian framework (i.e. teleportation of Gaussian states using Gaussian resources). The transformation of the Gaussian input state through the teleportation channel can be calculated by adding the first moments and CM of the channel to the first moments and CM of the initial state. In the limiting case of a perfect teleportation both \(\sqrt{2} m_f\) and \(2 \gamma_f\) are zero and the initial state is retrieved.

\subsection{Information theoretical characterization of CV quantum teleportation}\label{sec:information_theoretical_characterization_of_CV_quantum_teleportation_single_mode}
There exist different figures of merit to quantify the accuracy of the CV teleportation. Among others there is the fidelity of the quantum teleportation, detailing how closely the state arriving at  Bob's side resembles the original state from Alice. Another example is the classical communication capacity, giving the amount of classical information that can be pushed through the teleportation channel. In general, the choice of a figure of merit is motivated by its operational meaning.

In the scope of this paper, we characterize the teleportation channel in terms of its quantum capacity \cite{devetak_private_2005, lloyd_capacity_1997}, this means the highest rate at which quantum  information can be reliably transmitted through the channel when Alice and Bob make use of error correction to convey quantum information through the noisy channel. This choice seems to be the most natural  and appropriate, if quantum teleportation should be used to establish a true quantum link. 

For comparison purposes, we consider the \textit{two-way distillable entanglement} as another figure of merit in \ref{2waycc}. In this scenario, Alice and Bob also exchange classical information in a two-way fashion to extract maximally entangled states. In the main part of the paper, however, we will not allow two-way classical communication between Alice and Bob, because this approach delivers tighter bounds on the properties of the required resources. 

Indeed, the thermal-like noise added by the non-ideal teleportation can be counteracted by employing quantum error correction codes. These can increase the quality of the communication (e.g. in terms of the fidelity) at the cost of reducing the communication rate. The highest rate of reliable quantum communication, i.e. allowing asymptotically unit fidelity, is by definition the quantum capacity of the teleportation channel. The quantum capacity of Gaussian channels has been widely studied and characterized from an information theoretical perspective \cite{holevo_evaluating_2001, wolf_quantum_2007}. In full generality, the quantum capacity of a quantum channel $\phi$ is given by the following expression \cite{lloyd_capacity_1997, devetak_private_2005}:
\begin{equation}\label{Q}
Q = \max \left\{ 0, \lim_{\ell\to\infty} \frac{1}{\ell} \sup_{\hat\rho} I(\phi^{\otimes \ell},\hat\rho) \right\} \, ,
\end{equation}
where $\phi^{\otimes \ell}$ indicates $\ell$ parallel uses of the quantum channel. The entropic function
\begin{equation}
    I(\phi^{\otimes \ell},\hat\rho) = S[\phi^{\otimes \ell}(\hat\rho)] - S[(\phi^{\otimes \ell}\otimes\mathrm{id}_C)(|\psi\rangle_\rho\langle\psi|)] \, ,
    \label{eq:quantum_capacity}
\end{equation}
is known as the coherent information. Here, $S$ denotes the von Neumann entropy, $S[\hat\rho]=-\mathrm{Tr}(\hat\rho \ln{\hat\rho})$ (measured in {\it q-nats}\footnote{In order to obtain compact formulas for the quantum channel capacity bounds, we use natural logarithms, \(\ln = \log_e\).}). $|\psi\rangle_\rho$ is a purification of $\hat\rho$, involving an auxiliary quantum system denoted $C$, and $\mathrm{id}_C$ is the identity quantum channel acting on $C$. In general, it is very hard to evaluate the quantum capacity of a given channel, because one has to optimize \eref{eq:quantum_capacity} over all possible input states \(\hat\rho\) in the limit of infinite uses of the channel \(\phi\). An analytic formula for the quantum capacity is only known for few specimens of CV quantum channels \cite{wolf_quantum_2007}. It is however possible to evaluate upper and lower bounds of the quantum channel capacity.

In the following we put
\begin{eqnarray}\label{th-CM_single_mode}
    2\gamma_f = \left(\begin{array}{cc}
    N & 0 \\
    0 & N 
    \end{array}\right) \, .
\end{eqnarray}
This thermal-like form for the channel CM is the relevant one in several cases, as for the finitely squeezed EPR states with and without losses, where the parameter \(N\) contains the entanglement properties of the resource state.

\subsubsection{Lower Bound}\label{subsubsec:LowerBound}
A lower bound on the quantum capacity can be obtained by restricting ourselves in (\ref{eq:quantum_capacity}) to maximizing over Gaussian states $\hat\rho_G$, and by considering only a "single-use" of the channel, i.e. 
\begin{equation}\label{QGdef}
Q \geqslant \max \left\{ 0 , \sup_{\hat\rho_G} I(\phi,\hat\rho_G) \right\} =: Q_G \, .
\end{equation}
Clearly, a lower bound on the quantum capacity still provides an achievable rate of reliable communication\footnote{For the case of Gaussian channel, a natural conjecture is that Gaussian states saturate the maximization in \eref{Q}. However, it is in principle possible that the coherent information has a global maximum on non-Gaussian states. Moreover, as the coherent information might be super-additive for parallel channels, the regularized limit over $n$ is in general necessary for computing the quantum capacity \cite{smith_quantum_2011}.}. This lower bound can be computed efficiently for Gaussian channels \cite{holevo_evaluating_2001}. For the teleportation channel, it is a function of the noise CM in \eref{n-CM}. For a thermal-like noise with CM (\ref{th-CM_single_mode}), such a quantity was computed in \cite{holevo_evaluating_2001}, yielding:
\begin{equation}
    Q_G = \max \{ 0 , - 1 - \ln{N} \} \, .
    \label{eq:qg_bound_single_mode}
\end{equation}
The derivation of \eref{eq:qg_bound_single_mode}  is presented in \ref{app:QG}.

\subsubsection{Upper Bound}\label{subsubsec:UpperBound}
An upper bound on the quantum capacity can be calculated by noting that the thermal-like noise with CM (\ref{th-CM_single_mode}), for $N \leqslant 1$, can be simulated by the action of a linear amplifier with amplification factors $1/\eta$, followed by a linear attenuating channel with attenuating factor $\eta$. In fact, the composition of these channels transforms the input CM $\gamma_\rho$ to 
\begin{eqnarray}
    \gamma_\rho + \left(\begin{array}{cc}
    1-\eta & 0 \\
    0 & 1-\eta 
    \end{array}\right) \, ,
\end{eqnarray}
which coincides with the thermal-like channel by setting $\eta = 1 - N$. Due to the fact that the composition of channels cannot increase the quantum capacity, the capacity of the thermal-like channel is upper bounded by that of the attenuating channel. 

Using the results of \cite{wolf_quantum_2007} we can write
\begin{equation}
    Q \leqslant  \max \{ 0 , \ln{(1-N)}-\ln{N} \} =: Q_A \, .
    \label{eq:qa_bound_single_mode}
\end{equation}

\section{Single-mode quantum channel capacity analysis}\label{sec:single-mode_teleportation_analysis}
With formulas \eref{eq:qg_bound_single_mode} and \eref{eq:qa_bound_single_mode}, we are now able to evaluate bounds on the available quantum channel capacities of the standard one-mode quantum teleportation protocol.

At first, we assume that the shared bipartite entangled state is a finitely squeezed EPR state,
\begin{equation}
\ket{\psi}_{PDC} = \exp\left[r \left(\hat{a}^\dagger \hat{b}^\dagger - \hat{a} \hat{b} \right)\right] \ket{0} \, ,
\end{equation}
where the parameter $r$ describes the generated squeezing amplitude (we assume without loss of generality $r \geqslant 0$), which can be transformed into the squeezing value by the relation: \(\mathrm{squeezing}\mathrm[dB] = -10 \log_{10} \left( e^{-2 r}\right)\). Secondly, we study the effect of losses in the quantum capacity of the teleportation channel by assuming that the modes \(\left\{\hat{a}, \hat{a}^\dagger\right\}\), \(\left\{\hat{b}, \hat{b}^\dagger\right\}\) are attenuated by a factor $\eta$.

\subsection{Quantum channel capacity without losses}
If we neglect losses, which can occur during the EPR state distribution to Alice and Bob, the parameter in the CM \eref{th-CM_single_mode} reads $N = e^{-2r}$, where \(r\) labels the squeezing amplitudes of the shared EPR state. The bounds on the quantum channel capacities in \eref{eq:qg_bound_single_mode} and \eref{eq:qa_bound_single_mode} evaluate to the expressions:
\begin{eqnarray}
    Q_G & = & \max\{ 0 , 2r - 1 \} \,,
    \label{eq:qg_bound_sm_epr_state}\\
    Q_A & = & \max\{ 0 , 2r + \ln{(1-e^{-2r})} \} \, . 
    \label{eq:qa_bound_sm_epr_state} 
\end{eqnarray}
The limiting factor in the CV teleportation protocol is that EPR sources are constrained by the maximum amount of entanglement, and hence energy, that they are able to emit. For the case of PDC processes, this is equivalent to the overall optical gain of the down-conversion process. Furthermore, the channels used to transmit the EPR states to Alice and Bob are constrained by the amount of energy, that they can carry. For example, in the case of the ubiquitous optical fibers, the most prevalent method for quantum state distribution, transmitted pulses exceeding a certain power level undergo nonlinear optical processes in the fiber and subsequently lose part of their entanglement. 

\begin{figure}[htb]
    \begin{center}
        \includegraphics[width=0.6\textwidth]{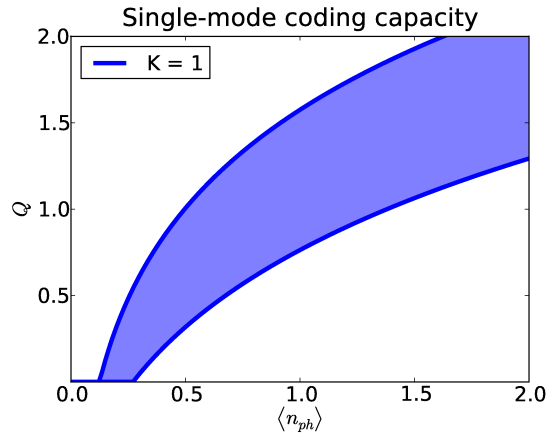}
    \end{center}
    \caption{Upper \(Q_A\) and lower \(Q_G\) bounds for the quantum channel capacity (measured in {\it q-nats}) of CV quantum teleportation using a single-mode EPR state. The minimum squeezing required in order to reliably transmit quantum information resides between 3.01 dB and 4.34 dB. (K=1: A single EPR state is transmitted.)}
    \label{fig:single_channel_capacity}
\end{figure}

It is hence vital to develop quantum communication protocols that encode quantum information as energy efficiently as possible. For this purpose, we benchmark quantum communication by evaluating the quantum channel capacity as a function of the energy, i.e. mean photon number \(\langle n_{ph} \rangle\) inside the channel. In the case of an EPR state this mean photon number is given as
\begin{equation}
    \left<n_{ph}\right> = \sinh^2(r) \, .
    \label{eq:mean_photon_number_single_mode}
\end{equation}
Figure \ref{fig:single_channel_capacity} displays the calculated upper and lower bounds \(Q_A\) and \(Q_G\), as defined in \eref{eq:qg_bound_sm_epr_state} and \eref{eq:qa_bound_sm_epr_state} as a function of the mean photon number \(\langle n_{ph} \rangle\) inside of the channel. 

This figure shows the minimum requirements for an EPR state to enable reliable quantum information transfer of the teleportation channel. The upper bound \(Q_A\) remains zero up to mean photon numbers \(\left<n_{ph}\right> = 0.125\) corresponding to squeezing values of 3.01 dB, whereas the lower bound \(Q_G\) is zero up to \(\left<n_{ph}\right> \approx 0.27\) equivalent to 4.34 dB of squeezing. Hence the minimum squeezing in EPR state allowing reliable quantum information transfer resides in the range between 3.01 dB and 4.34 dB. The situation changes if additional resources --- like unbounded two-way classical communication --- are allowed (see discussion in \ref{2waycc}).

\subsection{Quantum channel capacity including losses}
\begin{figure}[htp]
    \begin{center}
        \includegraphics[width=0.7\textwidth]{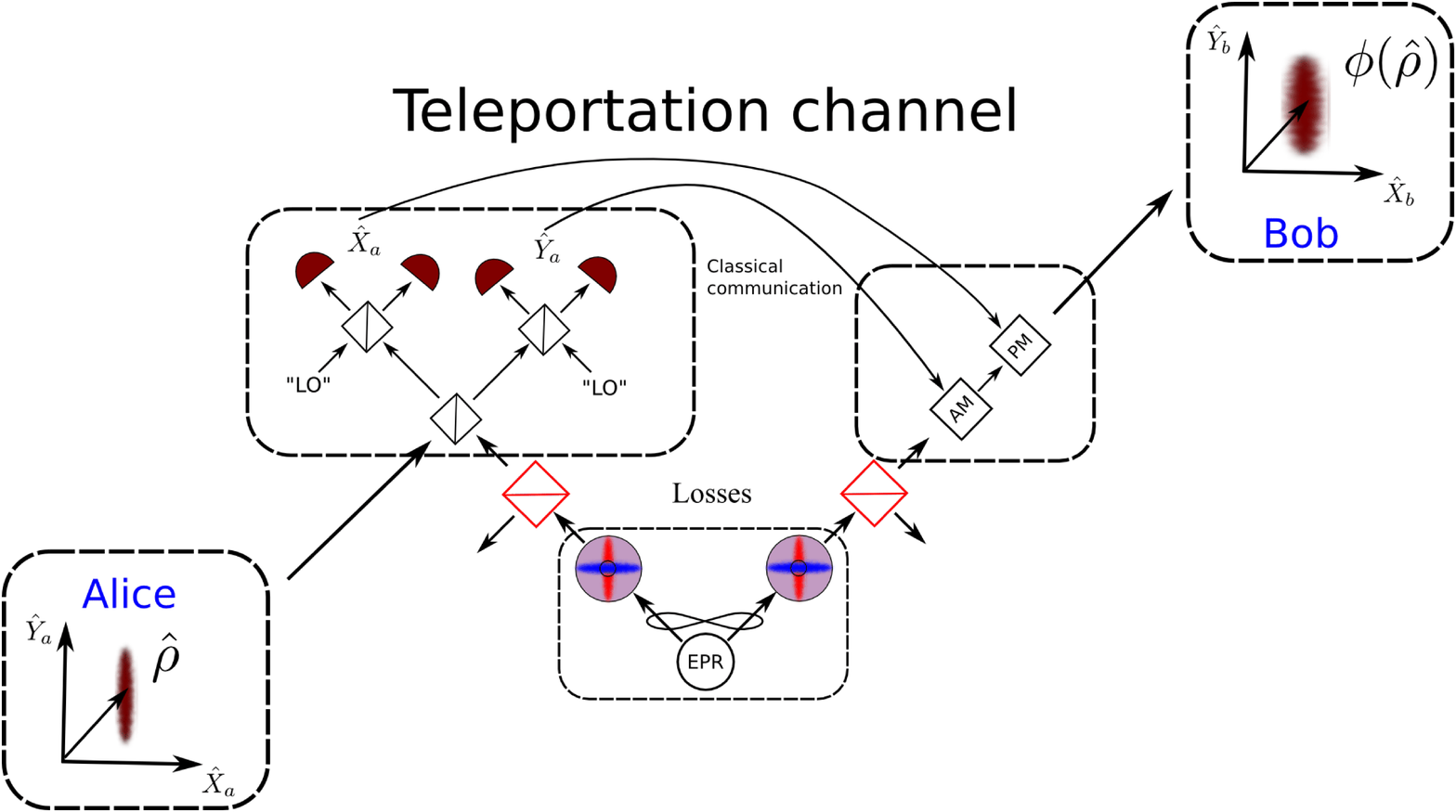}
    \end{center}
    \caption{CV teleportation setup including standard beam-splitter like losses during the distribution of the EPR state to Alice and Bob.}
    \label{fig:teleportation_with_losses_single_mode}
\end{figure}

Analyzing quantum teleportation in the framework of realistic applications, for example the ubiquitous quantum state \(\hat{\chi}\) distribution through optical fibers, the impact of losses has to be considered. We model these losses by the standard beam splitter interactions, \(\hat{a}\rightarrow \sqrt{\eta}\, \hat{a} + \sqrt{1-\eta} \, \hat{v}_a\), \(\hat{b}\rightarrow \sqrt{\eta}\, \hat{b} + \sqrt{1-\eta} \, \hat{v}_b\) during the distribution of the state to Alice and Bob, as displayed in figure \ref{fig:teleportation_with_losses_single_mode}, and evaluate the robustness of the state distribution as a function of the transmissivity of the channel \(\eta\).
With these conditions $N=\eta e^{-2r} + (1-\eta)$, and the channel capacity formulas evaluate to:
\begin{eqnarray}
Q_G & = & \max\{ 0 , - 1 - \ln{[1 - \eta (1-e^{-2r})]}\} \, ,
\label{eq:qg_bound_loss_single_mode}
\\
Q_A & = & \max\{ 0 , \ln{[\eta (1-e^{-2r})] - \ln{[1 - \eta (1-e^{-2r})]}} \} \, . 
\label{eq:qa_bound_loss_single_mode} 
\end{eqnarray}
\begin{figure}[htb]
    \begin{center}
        \includegraphics[width=0.6\textwidth]{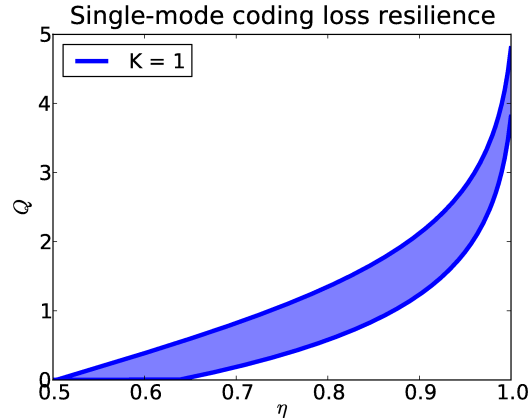}
    \end{center}
    \caption{Upper \(Q_A\) and lower \(Q_G\) bounds for the quantum channel capacity (measured in \textit{q-nats}) as a function of the transmissivity \(\eta\) for CV quantum teleportation using a single-mode EPR state including loss. The quantum channel capacity quickly degrades under loss until it reaches zero at loss rates exceeding 50\%. (K=1: A single EPR state is transmitted.)}
    \label{fig:single-mode_bounds_with_losses}
\end{figure}
Figure \ref{fig:single-mode_bounds_with_losses} depicts the quantum channel capacity as a function of the transmissivity \(\eta\) for an EPR state with a mean photon number of \(\langle n_{ph} \rangle = 30\).

Starting from a quantum channel capacity between 4 and 5 \textit{q-nats} it quickly degrades for decreasing transmissivities \(\eta\) until it reaches 0 at loss rates exceeding 50\%.

\section{Multi-mode EPR state generation and teleportation}\label{sec:mm_eps_state_generation_and_teleportation}
Having reviewed and established CV teleportation and the corresponding quantum communication rates in the single-mode regime we now expand the protocol to incorporate multiplexing. 

As discussed in the introduction there exist a variety of sources to create multi-mode EPR states. In the scope of this paper, we will focus on the properties of parametric down-conversion as a source of pulsed multi-mode EPR states in ultrafast frequency modes \cite{christ_probing_2011,  eckstein_highly_2011}. Yet our findings could also be adapted to other methods of squeezer generation as well.
\begin{figure}[htp]
    \begin{center}
        \includegraphics[width=\textwidth]{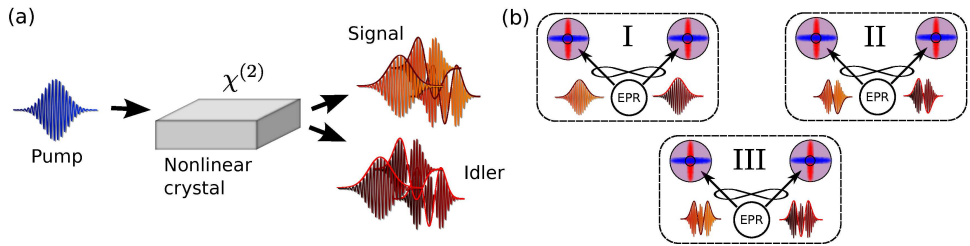}
    \end{center}
    \caption{(a) Schematic PDC process: An incoming ultrafast pump pulse is down-converted into two squeezed and entangled signal and idler waves. (b) The generated PDC state incorporates a multitude of EPR states in orthogonal ultrafast pulse modes.}
    \label{fig:multi-mode_epr_state_generation}
\end{figure}

Figure \ref{fig:multi-mode_epr_state_generation}(a) sketches the state generation process. An incoming ultrafast pump pulse decays inside a medium with a \(\chi^{(2)}\)-nonlinearity into two beams usually labelled signal and idler, which represent the two modes of the generated finitely squeezed EPR state. These states are well suited for quantum teleportation as they enable high repetition rates due to the ultrafast nature of the created pulses.

However, this PDC process pumped by a pulsed laser system produces not only a single EPR state but, as sketched in figure \ref{fig:multi-mode_epr_state_generation}(a) and (b), a multitude of ultrafast finitely squeezed EPR states into broadband frequency pulse modes.  Each output pulse consists of a multitude of EPR states in different orthogonal modes \cite{botero_modewise_2003, giedke_entanglement_2003}, formally described as
\begin{eqnarray}
    \ket{\psi}_{PDC} = \bigotimes_{k=1}^n \exp\left[r_k \left(\hat{A}_k^\dagger \hat{B}_k^\dagger - \hat{A}_k \hat{B}_k \right)\right] \ket{0},
    \label{eq:mm_epr_state}
\end{eqnarray}
where \(\hat{A}_k\) and \(\hat{B}_k\) label the different ultrafast pulse modes in the signal and idler arms, and the parameters \(r_k \geq 0 \) describe the generated squeezing amplitudes. A detailed derivation of \eref{eq:mm_epr_state} is given in \cite{christ_probing_2011}. For common PDC sources the squeezing parameters \(r_k\) form an exponentially decaying distribution, which can be engineered from emitting a single EPR state to creating a whole array of twin-beam squeezed states (see \cite{uren_photon_2003}).
\begin{figure}[htp]
    \begin{center}
        \includegraphics[width=\textwidth]{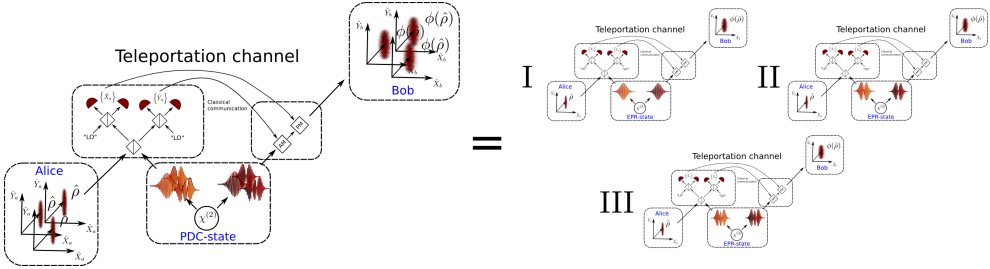}
    \end{center}
    \caption{Performing quantum teleportation using multi-mode PDC states in conjunction with multi-mode detection and displacements on Alice and Bob's side, effectively multiplexes the teleportation protocol.}
    \label{fig:multi-mode_teleportation}
\end{figure}

The standard protocol for single-mode CV teleportation \cite{braunstein_teleportation_1998} requires CV Bell-measurements, one-way classical communication and local phase-space displacements. In order to multiplex the teleportation protocol, these operations have to be performed on several pulse modes in parallel. There is a certain arbitrariness in that, because in principle different multi-mode orthogonal basis sets can be chosen for the quantum information encoding by the communicating parties Alice and Bob. However, in the following we are focusing on broadband entangled states produced via PDC, for which a unique natural mode basis \(\hat{A}_k\) \(\hat{B}_k\) arises from the Schmidt decomposition as given in \eref{eq:mm_epr_state}. In this basis each pair of modes \(\hat{A}_k\) and \(\hat{B}_k\) forms a finitely squeezed EPR state and we hence can treat each teleportation independently from the others. One could in principle also perform the teleportation in a different basis; this however would lead to correlations between all individual modes, reduce the individual mode entanglement and consequently lower the overall quality of the teleportation. It is hence natural to conjecture that the basis of the Schmidt modes optimizes the teleportation capacity. A detailed discussion of this issue will be presented elsewhere \cite{christ_preparation_2012}. 

These multi-mode PDC states are hence optimally suited to multiplex CV quantum teleportation as a single source is sufficient for creating many EPR states in multiple orthogonal ultrafast frequency modes. The general multiplexed protocol is depicted in figure \ref{fig:multi-mode_teleportation}. From the source a multitude of EPR states is transmitted to Alice and Bob. Alice now encodes the state she wants to teleport in the \(\{ \hat{A}_k\}\) modes of the source, superimposes the two beams at a beam splitter and then measures all optical modes separately. This can be implemented by either splitting the frequency modes into different spatial modes \cite{eckstein_quantum_2011, brecht_quantum_2011, raymer_interference_2010, mcguinness_quantum_2010} and guiding the light to independent measurement setups or by performing multi-mode homodyne detection \cite{beck_joint_2001, armstrong_programmable_2012}. These measurement results are then transmitted to Bob who performs the according displacements on each individual \(\hat{B}_k\) mode. He then retrieves the teleported multi-mode state \(\hat{\rho}_{tel}\).

The experimental implementation of multi-mode teleportation represents the main challenge for a deployment of our multi-mode coding protocol. Alice has to implement homodyne measurements in multiple orthogonal modes simultaneously on exactly the same basis as imposed by the multi-mode EPR source. Furthermore, the phase reference of the local oscillator beams has to be kept stable over all optical modes. Any errors in the measurement basis or phase mismatch between the individual modes will decrease the quantum communication rate. The same reasoning also applies to Bob who has to perform phase-locked displacements in the exact same basis. Although experimentally challenging, this problem is already addressed by various researchers working on multi-mode homodyne detection \cite{beck_joint_2001, armstrong_programmable_2012} and quantum pulse gates \cite{eckstein_quantum_2011, brecht_quantum_2011, raymer_interference_2010, mcguinness_quantum_2010}.

Eventually this approach of expanding the EPR-source and the detection apparatus to incorporate multiple modes allows us to perform multiplexed quantum teleportation. This in turn leads to several independent CV teleportation protocols being performed simultaneously.

\section{Multiplexed quantum channel capacity analysis}\label{sec:quantum_channel_capacity_calculations}
In this section, we characterize the multiplexed CV teleportation channel in terms of its quantum capacity. 

We consider two remarkable settings. Firstly, we assume that the teleportation resource is given by the multi-mode EPR state in equation (\ref{eq:mm_epr_state}). Secondly, we introduce a loss model in which each Schmidt mode is independently (and identically) attenuated by a standard beam splitter interaction with attenuation parameter $\eta$.

In both cases, the resulting multi-mode teleportation channel coincides with $n$ parallel single-mode teleportations. Hence, proceeding as in section \ref{subsubsec:LowerBound} and \ref{subsubsec:UpperBound} we obtain the lower bound on the multiplexed quantum channel capacity
\begin{equation}
    Q_G = \sum_{k=1}^n \max\{ 0 , - 1 - \ln{N_k} \} \, ,
    \label{eq:qg_bound}
\end{equation}
and the upper bound on the multiplexed quantum channel capacity
\begin{equation}
    Q_A = \sum_{k=1}^n \max \{ 0 , \ln{(1-N_k)}-\ln{N_k} \} \, , 
    \label{eq:qa_bound}
\end{equation}
for suitable parameters $N_k \geq 0$.

\subsection{Multi-mode teleportation}\label{sec:multi-mode_teleportation_analysis}
Neglecting losses during the EPR state distribution to Alice and Bob, the parameters $N_k$ are given by $N_k = e^{-2r_k}$, where \(r_k\) labels the individual squeezing amplitudes of the multi-mode squeezed state in \eref{eq:mm_epr_state}. The bounds on the quantum channel capacities in \eref{eq:qg_bound} and \eref{eq:qa_bound} evaluate to the straightforward expressions:
\begin{eqnarray}
    Q_G & = & \sum_{k=1}^n \max\{ 0 , 2r_k - 1 \} \,,
    \label{eq:qg_bound_mm_epr_state}\\
    Q_A & = & \sum_{k=1}^n \max\{ 0 , 2r_k + \ln{(1-e^{-2r_k})} \} \, . 
    \label{eq:qa_bound_mm_epr_state} 
\end{eqnarray}
The amount of energy of the multi-mode EPR state arriving at either Alice or Bob's side is related to the mean photon number in each arm given by:
\begin{equation}
    \left<n_{ph}\right> = \sum_k \sinh^2(r_k) \, .
    \label{eq:mean_photon_number}
\end{equation}
In analogy to the single-mode case, we analyze the teleportation channel as a function of the corresponding energy that now is expressed by the mean photon number \(\langle n_{ph} \rangle\) of all the modes involved in the teleportation protocol.

\begin{figure}[htp]
    \begin{center}
        \includegraphics[width=1.0\textwidth]{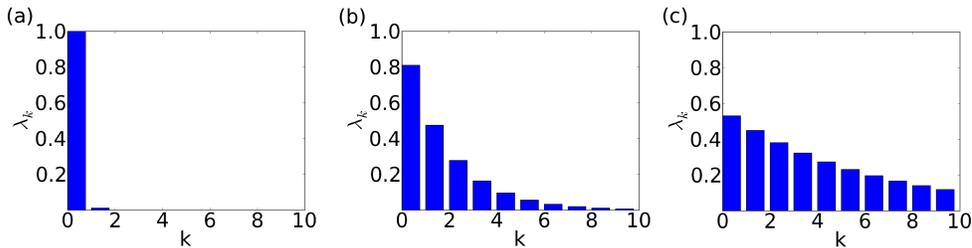}
    \end{center}
    \caption{Three different squeezer distributions \(\lambda_k\) normalized via \(\sum_k \lambda_k^2 = 1\) with varying degrees of multi-modeness. Depending on the source properties states ranging from a single squeezer (a) up to a whole range of EPR states in orthogonal optical modes are generated. Here \(k\) labels the number of the generated finitely squeezed EPR state and \(\lambda_k\) its amplitude relative to the other modes. \(\lambda_k\) can be converted to the actual squeezing amplitudes \(r_k\) via the overall optical gain \(B\) of the source: \(r_k = B \lambda_k\).}
    \label{fig:lambda_k_distribution_main}
\end{figure}

In order to compare the standard single-mode teleportation with our multiplexed coding, we simulated a PDC source creating EPR states multi-mode in frequency, based on the source employed in \cite{eckstein_highly_2011}. The source is able to operate in various degrees of multi-modeness and is hence perfectly suited for comparison purposes. We designed it to produce three different PDC states with varying numbers of modes as presented in figure \ref{fig:lambda_k_distribution_main}, which shows the three normalized exponentially decaying mode distributions and their different weights, which we use for this analysis.
These normalized mode distributions can be directly converted to the corresponding EPR state distributions, by multiplying them with the overall optical gain \(B\) of the process \(r_k = B \lambda_k\) (see \cite{eckstein_highly_2011} for details on the PDC source and \cite{christ_probing_2011}). We first simulate a purely single-mode source (figure \ref{fig:lambda_k_distribution_main}(a)), which only emits a single EPR state recreating the single-mode communication discussed in section \ref{sec:single-mode_teleportation_analysis} \cite{braunstein_teleportation_1998}. Figures \ref{fig:lambda_k_distribution_main}(b) and (c) present states with rising multi-mode character, many EPR states  generated in orthogonal pulse modes. Their \textit{effective} mode numbers \(K = 1 / \sum_k \lambda_k^4\) \cite{eberly_schmidt_2006}  are \(K = 1, 2\) and 6, where it should be stressed that, due to the generation process, not all modes share the same squeezing, but the entanglement follows an exponential decay towards higher-order modes.

\begin{figure}[htb]
    \begin{center}
        \includegraphics[width=\textwidth]{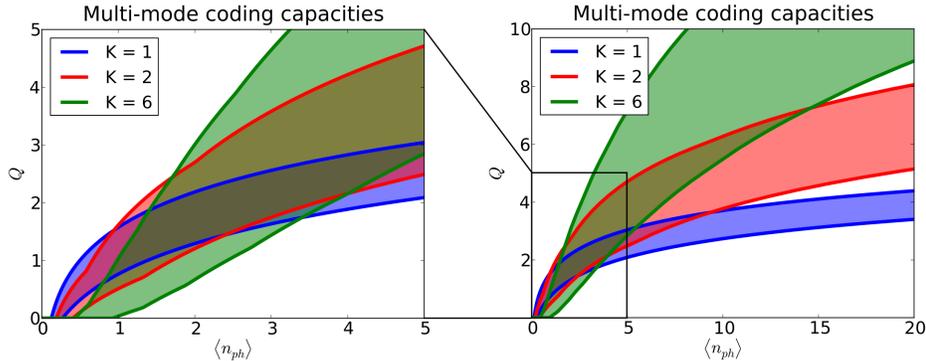}
    \end{center}
    \caption{Quantum channel capacity bounds (in {\it q-nats}) for multi-mode transmission. From bottom to top $K=1,2,6$. Applying multi-mode EPR states for the teleportation gives a significant increase in the available quantum channel capacity as soon as a certain energy threshold is exceeded. This is due to the increased energy efficiency of multi-mode coding in conjunction with the fact that a minimum amount of squeezing has to be present in each optical mode to achieve positive quantum channel capacities (see section \ref{sec:single-mode_teleportation_analysis}).}
    \label{fig:multi-mode_bounds}
\end{figure}

Using \eref{eq:qg_bound_mm_epr_state} and \eref{eq:qa_bound_mm_epr_state} we derive the lower and upper quantum channel capacity bounds \(Q_G\) and \(Q_A\) for the different squeezer distributions presented in figure \ref{fig:lambda_k_distribution_main}. The obtained quantities are plotted in figure \ref{fig:multi-mode_bounds} as function of the mean photon number or energy inside the channel.

It is evident that the multiplexed teleportation relying on several less squeezed optical modes results in significantly higher bounds on the channel capacities with respect to the standard single-mode coding as soon as a certain energy threshold is exceeded. While the blue shaded area, which corresponds to single-mode teleportation, with the complete energy being concentrated in a single mode, never reaches quantum channel capacities above 5 {\it q-nats} in the considered energy range, encoding information on multiple modes shows significantly higher quantum channel capacities\footnote{As an alternative to frequency multiplexing one could also transmit multiple weakly squeezed EPR states in succession instead of one strongly squeezed EPR state. Mathematically both approaches are equivalent.}. 

The underlying reason for this behaviour is the efficiency of the EPR state distribution. Following the discussion in \cite{christ_probing_2011} one finds that it is far more efficient, in terms of energy content, to utilize several EPR states with a low amount of squeezing than one EPR state with a high squeezing value. A similar effect is also observed in other contexts such as energy efficient entanglement creation \cite{van_enk_entangled_2005}, quantum reading \cite{pirandola_quantum_2011, pirandola_quantum_2011-1} and entanglement distribution \cite{kraus_discrete_2004}. However, the fact that a certain energy is required to achieve a positive quantum channel capacity (see section \ref{sec:single-mode_teleportation_analysis}) counteracts the enhanced energy efficiency of multi-mode coding and consequently there exists a trade-off between using as many optical modes as possible for enhanced energy efficiency and sufficiently few optical modes to achieve positive quantum channel capacities. 

\subsection{Optimal multi-mode coding}\label{sec:optimal_multi-mode_coding}
In order to achieve the optimal quantum channel capacity one has to carefully balance the splitting of the energy into different modes. As discussed in section \ref{sec:single-mode_teleportation_analysis} the upper bound \(Q_A\) will drop to zero as soon as the applied EPR state is below 3.01 dB. Hence, in order to maximize the quantum channel capacity of CV teleportation, one has to distribute the energy over as many EPR states as possible while the created EPR states still bear sufficiently high squeezing values. 

We analyzed the optimal number of modes for multiplexing that achieves maximal quantum channel capacities for a given amount of energy (mean photon number \(\left<n_{ph}\right>\)). Our following discussion of the encoding into the optimal number of modes is split into two parts: First we will elaborate on PDC sources that can be realized in a straightforward manner by use of existing setups, and discuss their optimal design. Then we turn our attention to the global optimum where the necessary squeezer distributions would require further engineering of the source.

\subsubsection{Common EPR-sources}\label{sec:commmon_epr_sources}
\begin{figure}[htb]
    \begin{center}
        \includegraphics[width=\textwidth]{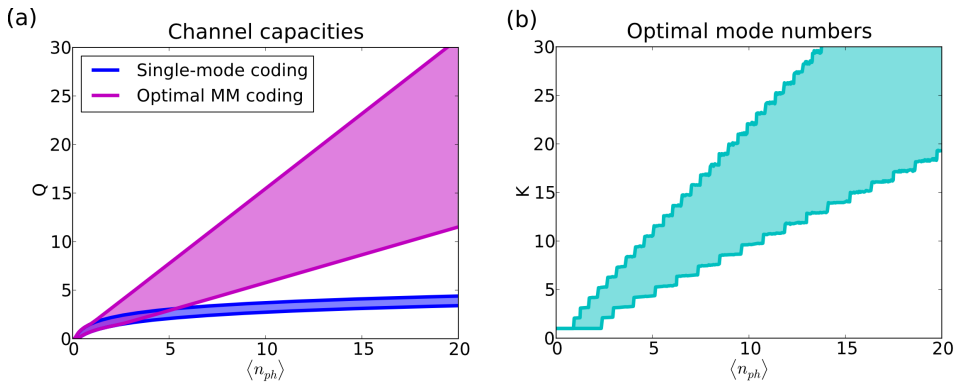}
    \end{center}
    \caption{(a) \(Q_A\) and \(Q_G\) channel capacities (in {\it q-nats}) for single-mode and optimal multi-mode coding given a \textit{common} EPR-source. (b) Effective mode number \(K\) required for the optimal multi-mode coding. Adapted multi-mode codes achieve quantum channel capacities outperforming single-mode approaches.}
    \label{fig:optimal_bounds}
\end{figure}

Given a common source of multi-mode EPR states --- as presented in \cite{eckstein_highly_2011} --- we optimize the capacities \(Q_A\) and \(Q_G\) over all possible \textit{effective} mode numbers \(K\) for each mean photon number \(\left<n_{ph}\right>\) under the restriction of a mode distributions \(r_k\) given by the formula \cite{uren_photon_2003}:
\begin{equation}
    r_k = B \sqrt{1-\mu^2} \mu^k \, , \qquad 0 \le \mu \le 1 \, .
    \label{eq:squeezer_distribution}
\end{equation}
The results are depicted in figure \ref{fig:optimal_bounds}.
Figure \ref{fig:optimal_bounds}(a) shows the \(Q_A\) and \(Q_G\) bounds for the standard  single-mode CV teleportation in comparison with the obtained optimized multi-mode coding. In the case of low energies both approaches yield identical rates. However given mean photon numbers above \(\left<n_{ph}\right> \approx 0.94\) (7.47 dB) and \(\left<n_{ph}\right> \approx 2.40\) (10.61 dB) for \(Q_A\) and \(Q_G\) respectively the optimized multi-mode coding outperforms the single-mode approach in each bound individually. Finally, the lower bound \(Q_G\) of the optimized multi-mode encoding surpasses the upper bound \(Q_A\) of single-mode coding at  \(\left<n_{ph}\right> \approx 5.37\) (13.70 dB). 

Most importantly, however, the optimal coding bounds show a linear increase in channel capacity with energy, whereas the single-mode quantum capacity bounds exhibit a logarithmic growth for high mean photon numbers. Consequently, multi-mode coding enables an \textit{exponential} increase of the quantum communication rate over single-mode coding in the presence of energy constraints. The effective mode number \(K\) corresponding to the optimal bounds in figure \ref{fig:optimal_bounds}(a) are presented in figure \ref{fig:optimal_bounds}(b). As the channel capacities they feature a (mostly) linear increase with energy.

\subsubsection{Optimal encoding with EPR-sources}\label{sec:optimal_epr_sources}
The main drawback of the currently available PDC sources emitting EPR states is that they feature exponentially decaying squeezing amplitudes \(r_k\) for higher-order modes, as already depicted in figure \ref{fig:lambda_k_distribution_main}. This is not the optimal encoding because a certain number of squeezers will always reside below the bound to create positive quantum channel capacities. Hence, they do not contribute to the quantum communication rate while still occupying energy. 

We can negate this drawback by applying multi-mode EPR states exhibiting a \textit{flat} distribution \(r_k = r\) with a mode number \(K\). Experimentally these states can be approximated by engineering the pump pulse and the phase-matching of the PDC process. This flat distribution offers the great advantage that all EPR states contribute to the overall channel capacity and no energy is lost in weakly squeezed modes with zero capacity. Indeed, it can be proven to provide the optimal distribution of the squeezing amplitudes, see \ref{Sec:OptDistr}.

In the optimal case of flat mode distributions the formulas for \(Q_G\) and \(Q_A\), as a function of the mode number \(K\) and mean photon number \(\langle n_{ph} \rangle\), evaluate to:
\begin{eqnarray}
    \label{eq:QG_flat_coding}
    \fl \qquad Q_G = \max \left\{0, K \left[2 \, \mathrm{arcsinh}\left(\sqrt{\frac{\langle n_{ph} \rangle}{K}}\right) -1 \right] \right\} \, ,  \\
    \nonumber
    \fl \qquad Q_A =  \max \left\{0, K \left[ 2  \, \mathrm{arcsinh}\left(\sqrt{\frac{\langle n_{ph} \rangle}{K}}\right) \right. \right.\\
    \qquad + \left. \left. \ln\left(1 - \exp\left(-2 \, \mathrm{arcsinh}\left(\sqrt{\frac{\langle n_{ph} \rangle}{K}}\right) \right)\right) \right] \right\} \, .
    \label{eq:QA_flat_coding}
\end{eqnarray}

\begin{figure}[htb]
    \begin{center}
        \includegraphics[width=\textwidth]{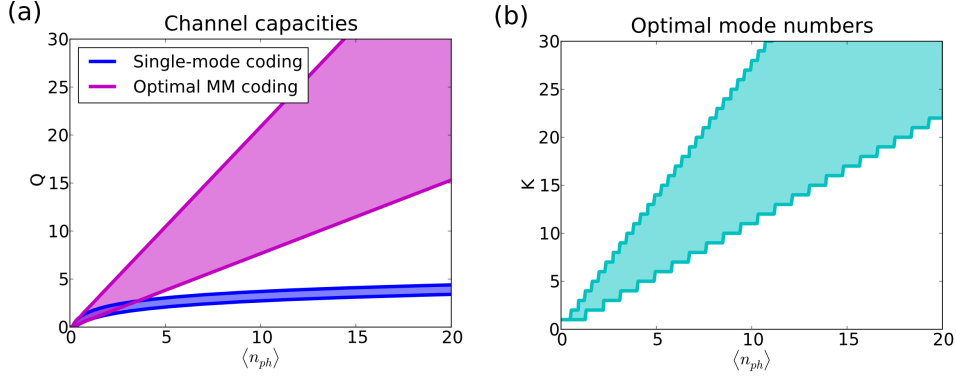}
    \end{center}
    \caption{(a) \(Q_A\) and \(Q_G\) channel capacities, measured in {\it q-nats}, for single-mode and optimal multi-mode coding given a \textit{flat} mode distribution. (b) Effective mode number \(K\) required for the optimal multi-mode coding. Adapted multi-mode codes achieve quantum channel capacities outperforming single-mode approaches.}
    \label{fig:optimal_bounds_flat}
\end{figure}

We analyze the achievable channel capacities in this optimized configuration by maximizing over the mode number \(K\) for given energies or mean photon numbers \(\langle n_{ph} \rangle\). The results are displayed in figure \ref{fig:optimal_bounds_flat}. Similar to the common EPR state distributions discussed in section \ref{sec:commmon_epr_sources} they feature the advantage of showing a linear gain with mean photon number \(\langle n_{ph} \rangle\) instead of the logarithmic growth present in the single-mode coding case and hence an \textit{exponential} growth in quantum communication rate. The achievable channel capacities surpass the quantum communication rates available using common EPR states as displayed in figure \ref{fig:optimal_bounds}, since no energy is located in weakly squeezed EPR states that do not contribute to the overall quantum channel capacity.

Furthermore \eref{eq:QG_flat_coding}, enables us to directly assess the optimal number of modes \(K_{opt}\) required to encode information for optimal capacity given a certain mean photon number \(\langle n_{ph} \rangle\):
\begin{eqnarray}
    K_{opt}(Q_G) \approx 1.1133 \, \langle n_{ph} \rangle \, , \qquad 
    K_{opt}(Q_A) \approx 2.7523 \, \langle n_{ph} \rangle \, .
    \label{eq:optimal_mode_number_flat_coding}
\end{eqnarray}
From equation \eref{eq:optimal_mode_number_flat_coding} we conclude that for the optimum mode number the squeezing of individual modes stays fixed between 4.96 dB and 7.33 dB. Consequently using energy to achieve squeezing values above this threshold is actually detrimental for the overall quantum capacity and it is much more resourceful employing it to create EPR states in additional modes.

\subsection{Multi-mode analysis under loss}\label{sec:multi-loss}
We finally consider the impact of loss for multi-mode coding similar to the single mode case discussed in section \ref{sec:single-mode_teleportation_analysis}. For a first analysis of the robustness under losses, we assume that all the modes are attenuated by the same attenuation factor $\eta$. The more realistic setting of frequency depending attenuation will be considered
elsewhere \cite{christ_preparation_2012}.
\begin{figure}[htp]
    \begin{center}
        \includegraphics[width=0.8\textwidth]{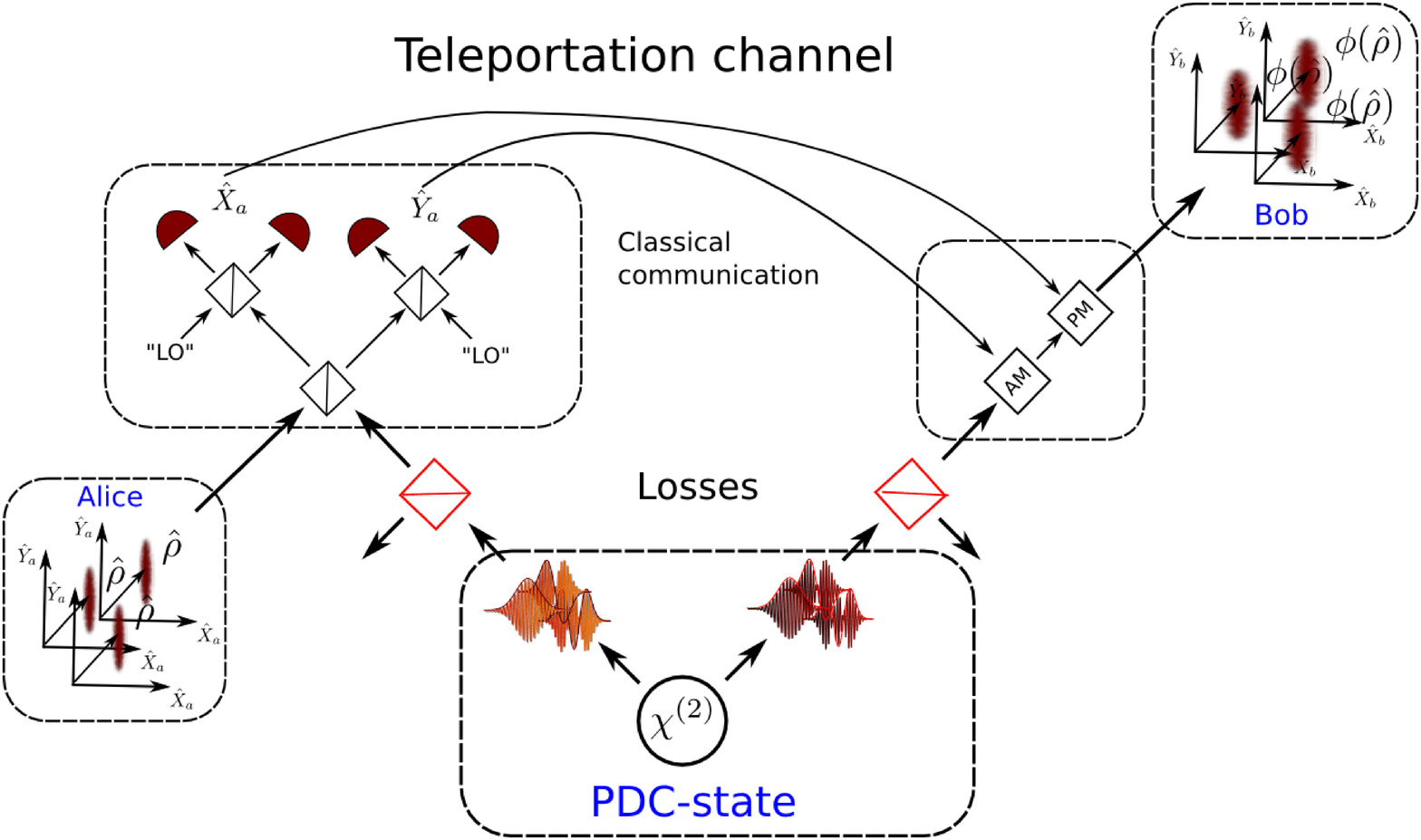}
    \end{center}
    \caption{Frequency multi-mode teleportation setup including standard beam-splitter like losses during the distribution of the EPR states to Alice and Bob.}
    \label{fig:teleportation_with_losses}
\end{figure}
Under these conditions the channel capacity formulas evaluate to:
\begin{eqnarray}
Q_G & = & \sum_{k=1}^n \max\{ 0 , - 1 - \ln{[1 - \eta (1-e^{-2r_k})]}\} \, ,
\label{eq:qg_bound_loss}
\\
Q_A & = & \sum_{k=1}^n \max\{ 0 , \ln{[\eta (1-e^{-2r_k})] - \ln{[1 - \eta (1-e^{-2r_k})]}} \} \, . 
\label{eq:qa_bound_loss} 
\end{eqnarray}

\begin{figure}[htb]
    \begin{center}
        \includegraphics[width=0.6\textwidth]{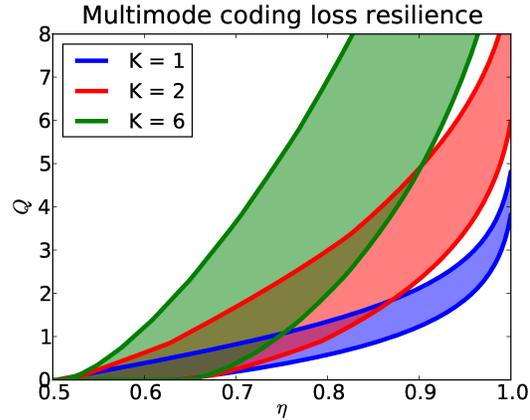}
    \end{center}
    \caption{The loss resilience of the quantum information transmission rate visualized for multi-mode and single-mode coding. From bottom to top $K=1,2,6$. Multi-mode coding offers the advantage of an increased loss resilience and gives significant higher rates over almost the whole \(\eta\) range in comparison to the single-mode approach.} 
    \label{fig:multi-mode_bounds_with_losses}
\end{figure}
Using \eref{eq:qg_bound_loss} and \eref{eq:qa_bound_loss} we determine the loss resilience of the three exemplary states. We start by tuning the three test states to exhibit identical mean photon numbers \(\left<n_{ph}\right> = 30\) and study their behaviour under loss. Our results are visualized in figure \ref{fig:multi-mode_bounds_with_losses} where we plot the quantum channel capacity as a function of the transmissivity \(\eta\). Clearly an enhanced loss resilience is observed for the multi-mode coding with respect to the single-mode protocol which quickly degenerates under loss. The reason for this advantage is well known: Strongly squeezed EPR states are highly susceptible to loss whereas the encoding of information on multiple weakly squeezed states is much more robust against this type of noise (see, e.g., \cite{van_enk_entangled_2005}).

\subsection{Optimal multi-mode coding under loss}
In a similar manner to the discussion in section \ref{sec:optimal_multi-mode_coding} we search for the optimal number of modes to encode information yet including loss during the EPR state transmission. 
\begin{figure}[htb]
    \begin{center}
        \includegraphics[width=\textwidth]{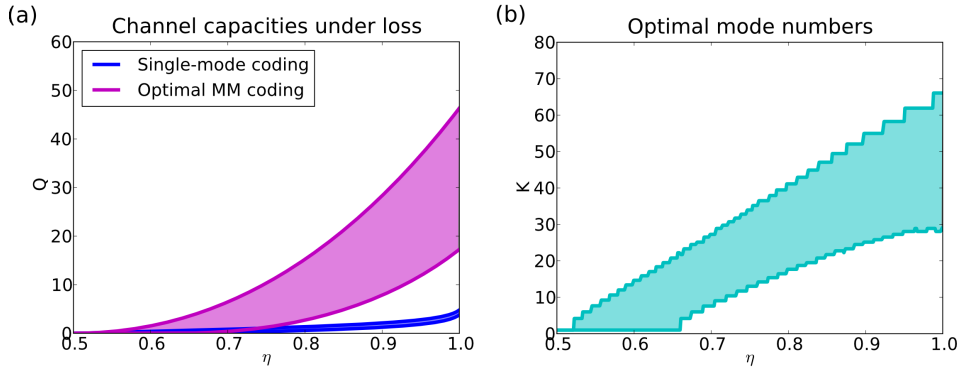}
    \end{center}
    \caption{(a) \(Q_A\) and \(Q_G\) channel capacities (measured in {\it q-nats}) for single-mode and optimal multi-mode coding given a \textit{common} mode distribution as a function of loss. (b) Effective mode number \(K\) required for the optimal multi-mode coding. Adapted multi-mode codes outperform single-mode approaches in the low-loss regime.}
    \label{fig:optimal_coding_w_loss}
\end{figure}

For this purpose we use an input state with mean photon number \(\langle n_{ph} \rangle = 30\) and in dependence of the transmissivity \(\eta\) optimize the channel capacity over all possible input mode distributions. In figure \ref{fig:optimal_coding_w_loss} (a), we display the achievable rates using common squeezer distributions readily available in the lab, as already discussed in section \ref{sec:commmon_epr_sources}. Figure  \ref{fig:optimal_coding_w_loss} (b) depicts the \textit{effective} mode numbers required to achieve the optimal coding. This analysis shows that in the case of losses the optimal squeezing values differ from the ones for lossless coding (see section \ref{sec:multi-loss}) and the advantages of multiplexing are partially lost depending on the amount of loss in the system. In the low-loss regime the optimized multi-mode coding outperforms the standard single-mode approach. However in the case of high losses approaching 50\% --- the exact value depends on the initial energy or mean photon number --- the single-mode coding surpasses our multi-mode approach. This is to be expected for the applied CV quantum communication protocol since it is not designed for transmission under extreme loss but for low-loss applications. Its optimal operational area is the transmission of large amounts of quantum information over short distances where it excels. For quantum communication over longer distances --- without repeater stations --- other quantum communication protocols are more suitable.

\begin{figure}[htb]
    \begin{center}
        \includegraphics[width=\textwidth]{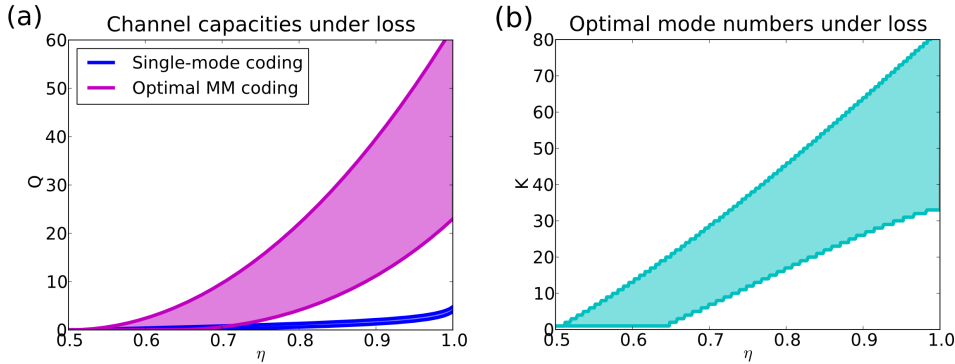}
    \end{center}
    \caption{(a) \(Q_A\) and \(Q_G\) channel capacities (in {\it q-nats}) for single-mode and optimal multi-mode coding given an optimal \textit{flat} mode distribution as a function of loss. (b) Effective mode number \(K\) required for the optimal multi-mode coding. Adapted multi-mode codes outperform single-mode approaches in the low-loss regime.}
    \label{fig:optimal_coding_flat_modes_w_loss}
\end{figure}
However, these results are still not optimal. For this purpose, we investigated the attainable quantum channel capacities using a flat mode distributions as discussed in section \ref{sec:optimal_epr_sources}. The attainable rates are presented in figure \ref{fig:optimal_coding_flat_modes_w_loss} (a) and (b). Again the optimized coding on flat mode distributions outperforms the single-mode coding in the low-loss regime and achieves higher rates than the use of common squeezer distributions.

\begin{figure}[htb]
    \begin{center}
        \includegraphics[width=\textwidth]{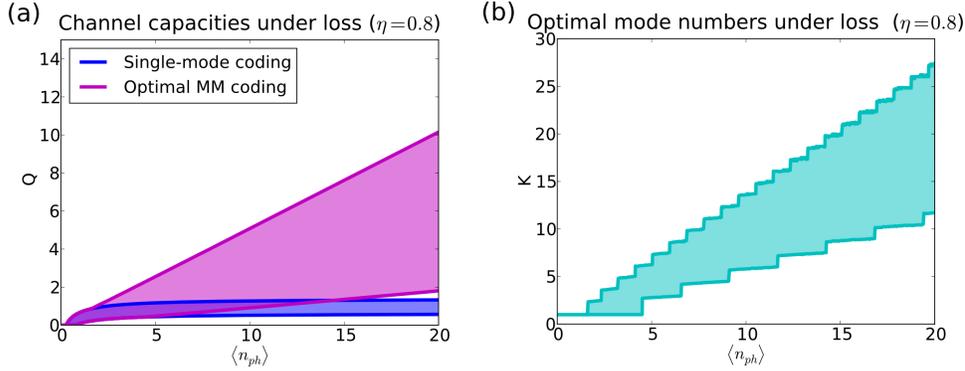}
    \end{center}
    \caption{(a) \(Q_A\) and \(Q_G\) (measured in \textit{q-nats}) for single-mode and optimal multi-mode coding given a \textit{common} mode distribution as a function of energy for a constant loss rate of \(\eta = 0.8\). (b) Effective mode number \(K\) required for the optimal multi-mode coding. Even when considering losses the multi-mode coding shows an linear increase with energy, which constitutes an exponential increase over the logarithmic growth of the single-mode protocol.}
    \label{fig:optimal_coding_exp_mode_w_loss_energy}
\end{figure}
\begin{figure}[htb]
    \begin{center}
        \includegraphics[width=\textwidth]{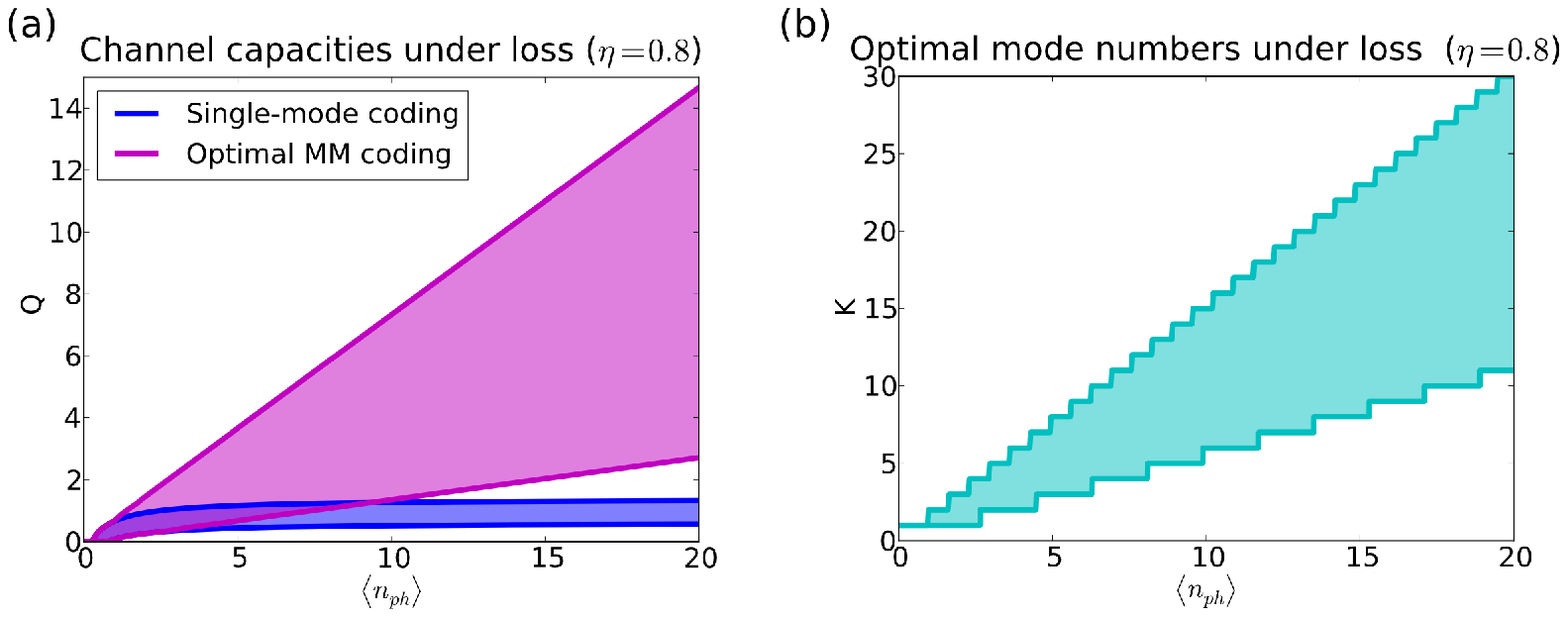}
    \end{center}
    \caption{(a) \(Q_A\) and \(Q_G\) (measured in \textit{q-nats} for single-mode and optimal multi-mode coding given a \textit{flat} mode distribution as a function of energy for a constant loss rate of \(\eta = 0.8\). (b) Effective mode number \(K\) required for the optimal mulit-mode coding. Even when considering losses the multi-mode coding shows an linear increase with energy, which constitutes an exponential increase over the logarithmic growth of the single-mode protocol.}
    \label{fig:optimal_coding_flat_mode_w_loss_energy}
\end{figure}
Next, we turn our attention to the quantum communication rates as a function of the energy for a constant loss rate. In figure \ref{fig:optimal_coding_exp_mode_w_loss_energy}(a), we plot the optimal multi-mode coding quantum channel capacities for a transmissivity of \(\eta = 0.8\) for \textit{common} squeezer distributions as a function of energy. The linear dependence of multi-mode quantum communication on energy for lossless coding (see section \ref{sec:optimal_multi-mode_coding}) remains in this setting including losses during the state transmission. The single-mode coding also still features a logarithmic growth as a function of energy similar to the one observed for lossless state transmission. Consequently, the multi-mode protocol achieves an \textit{exponential} increase over single-mode coding even in the presence of loss, as long as a certain minimum amount of energy is used in the communication. 

This effect is even more prominent when we consider optimal \textit{flat} multi-mode EPR state distributions, as depicted in figure \ref{fig:optimal_coding_flat_mode_w_loss_energy}(a). It achieves higher quantum communication rates in comparison to the multi-mode coding on \textit{common} squeezer distribution, while still featuring the linear growth as a function of energy as present in the lossless coding discussed in section \ref{sec:optimal_epr_sources}.

\begin{figure}[htb]
    \begin{center}
        \includegraphics[width=\textwidth]{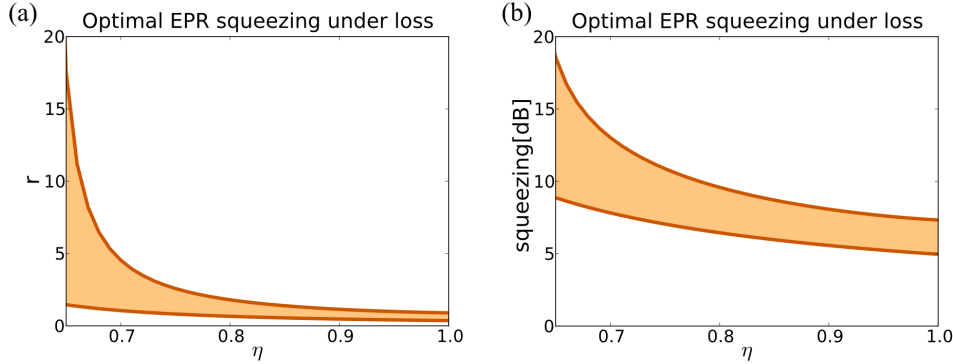}
    \end{center}
    \caption{The optimal squeezing values, in \(r\) (a) and dB (b), for the individual EPR states in the multiplexed protocol adapted to the losses in the channel.}
    \label{fig:optimal_squeezing_values_under_loss}
\end{figure}

However, to achieve the optimal quantum channel capacity, the squeezing values of the individual EPR states in the communication protocol have to be adapted to the losses in the channel. Starting from the aforementioned 4.96 dB and 7.33 dB discussed in section \ref{sec:optimal_epr_sources} for lossless communication, rising amounts of EPR squeezing are required for optimal coding. The exact values, as a function of the transmissivity \(\eta\), are depicted in figure \ref{fig:optimal_squeezing_values_under_loss}.

In summary, even in the presence of loss, multi-mode coding not only gives an \textit{exponential} increase in the observed quantum communication rate in comparison to single-mode coding as a function of energy, but also features an enhanced loss resilience.

\section{Conclusion}\label{sec:conclusion}
In conclusion, we expanded the theory of CV quantum teleportation into the multi-mode domain and presented a practical approach to implement the proposed multiplexing protocol. We calculated upper and lower bounds on the attainable quantum channel capacities by encoding information on multiple optical modes. Our analysis reveals that multiplexing not only features an enhanced energy efficiency leading to an \textit{exponential} increase in the achievable quantum communication rates in comparison to single-mode coding, but also gives an improved loss resilience. 

However, as reliable quantum information transfer is achieved only for squeezed modes above a certain threshold value, a careful optimization of the number of used coding modes is needed.

Our findings show that EPR states with squeezing values between 3.01 and 4.34 dB are required for having reliable quantum information transfer through the teleportation channel. Due to the energy constraints inside a quantum channel the optimum is reached when EPR states with squeezing values in the range from 4.96 dB up to 7.33 dB are employed. Creating squeezing above this bound is actually detrimental to the overall quantum communication rate. It is much more resourceful to invest the excess energy in creating EPR states in multiple optical modes.

\section{Acknowledgments}
The authors thank Benjamin Brecht for help with the PDC calculations. They also thank Stefano Mancini, Vittorio Giovannetti, Filippo Caruso, and Stefano Pirandola, for useful comments.

The research leading to these results has received funding from the European Commission's seventh Framework Programme CORNER (FP7/2007-2013) under grant agreement no.~213681.

\begin{appendix}

\section{Calculation of the lower bound $Q_G$}\label{app:QG}
For computing $Q_G$, we have to maximize the coherent information over Gaussian states. In this case, we can assume without loss of generality that \(|\psi\rangle_{\rho_G}\) is an EPR state with squeezing parameter $s$, shared between the subsystem $A$ and the auxiliary subsystem $C$, described by the Wigner function,
\begin{equation}
    W_{|\psi\rangle_{\rho_G}\langle\psi|}(q_A, p_A; q_C, p_C) = G_{(0,\gamma^{AC}_s)}(q_A, p_A; q_C, p_C) \, ,
\end{equation}
where
\begin{eqnarray}
    \gamma^{AC}_s = \frac{1}{2}
    \left( 
    \begin{array}{cccc}
    \cosh{2s} & 0          & \sinh{2s} & 0            \\
    0         & \cosh{2s}  & 0         & -\sinh{2s} \\
    \sinh{2s} & 0          & \cosh{2s} & 0            \\
    0         & -\sinh{2s} & 0         & \cosh{2s}
    \end{array}
    \right) \, .
\end{eqnarray}
The action of the channel transmitting the state of subsystem \(A\) from Alice to Bob, transforms this state to
\begin{equation}
    W_{(\phi\otimes\mathcal{I})(|\psi\rangle_{\rho_G}\langle\psi|)}(q_B, p_B; q_C, p_C) = G_{(0,\gamma^{BC}_s)}(q_B, p_B; q_C, p_C) \, ,
     \label{2squeezer}
\end{equation}
with
\begin{eqnarray}\label{2squeezerCM}
    \gamma^{BC}_s = \frac{1}{2}
    \left( 
    \begin{array}{cccc}
    2N+\cosh{2s} & 0           & \sinh{2s} & 0            \\
    0           & 2N+\cosh{2s} & 0         & -\sinh{2s} \\
    \sinh{2s}   & 0           & \cosh{2s} & 0            \\
    0           & -\sinh{2s}  & 0         & \cosh{2s}
    \end{array}
    \right) \, .
\end{eqnarray}
which is known as the Choi-Jamio\l{}kowski (CJ) state associated with the channel. After tracing out the \(C\) subsystem the reduced state of subsystem $B$ takes on the form
\begin{equation}
    W_{\phi(\rho)}(q_B, p_B) = G_{(0,\gamma^B_s)}(q_B, p_B) \, ,
    \label{eq:coherent_information_reduced_state}
\end{equation}
with
\begin{eqnarray}
    \gamma^B_s = \frac{1}{2}
    \left( 
    \begin{array}{cc}
    2N+\cosh{2s} & 0           \\
    0           & 2N+\cosh{2s} 
    \end{array}
    \right) \, .
\end{eqnarray}
In order to evaluate \(Q_G\), we have to determine the von Neumann entropy of the two states in \eref{2squeezer} and \eref{eq:coherent_information_reduced_state}. In the case of Gaussian states this is a straightforward calculation, because the state is defined by its CM the von Neumann entropy is determined by their symplectic eigenvalues \cite{holevo_evaluating_2001, ferraro_gaussian_2005}.
Then we have 
\begin{equation}
S[\phi(\rho)] = g( \nu^B - 1/2 ) \, ,
\end{equation}
where $g(w):= (w+1)\ln{(w+1)} - w\ln{w}$, and $\nu^B$ is the symplectic eigenvalue of the CM $\gamma^B_s$. 
The symplectic eigenvalue is calculated from the matrix $\Omega \gamma^B_s$, where $\Omega = \imath \sigma_2$ 
is the symplectic form, with
\begin{eqnarray}
    \imath\sigma_2 = \left(\begin{array}{cc}
    0 & -1 \\
    1 & 0 
    \end{array}\right) \, .
\end{eqnarray}
In particular, the eigenvalues of $\Omega \gamma^B_s$ are $\pm \imath \nu^B$.

Similarly, 
\begin{equation}
S[(\phi\otimes\mathcal{I})(|\psi\rangle_\rho\langle\psi|)] = g( \nu^{BC}_+ - 1/2 ) + g( \nu^{BC}_- - 1/2 ) \, ,
\end{equation}
where $\nu^{BC}_\pm$ are the symplectic eigenvalues of the CM $\gamma^{BC}_s$, where the $\pm \imath \nu^{BC}_+$ and $\pm \imath \nu^{BC}_-$ are the eigenvalues of $(\Omega\oplus\Omega)\gamma^{BC}_s$.

The resulting coherent information is an increasing function of $s$:
\begin{equation}
\nu^{B} = N + \frac{1}{2}\cosh{2s} \, ,
\end{equation}
\begin{equation}
\nu^{BC}_\pm = \frac{1}{2} \sqrt{ 1 + 2N^2 + 2N\cosh{2s} \pm 2N \sqrt{1+N^2+2N\cosh{2s}} } \, .
\end{equation}
In the limit of an infinitely squeezed state ($s \to \infty$), we obtain 
\begin{equation}
\nu^B \simeq N + \frac{1}{4}e^{2s} \, ,
\end{equation}
and
\begin{equation}
\nu^{BC}_\pm \simeq \frac{e^s \sqrt{N}}{2} \pm N \, .
\end{equation}
Finally, after straightforward algebra, we obtain
\begin{eqnarray}
Q_G & = & \max \left\{ 0 , \lim_{s\to\infty} g( \nu^B - 1/2 ) - g( \nu^{BC}_+ - 1/2 ) - g( \nu^{BC}_- - 1/2 ) \right\} \nonumber \\
    & = & \max \{ 0 , - 1 - \ln{N} \} \, .
\end{eqnarray}

\section{Classical communication allowed}\label{2waycc}
In the main part of the paper, we have considered a scenario in which Alice and Bob make use of error correction to convey quantum information through the noisy teleportation channel. Alternatively, if they are also allowed to exchange classical information in a two-way fashion, they can perform a protocol of entanglement purification to extract maximally entangled states up to a rate equal to the {\it two-way distillable entanglement} \cite{bennett_mixed-state_1996}, denoted $D_2$, of the CJ state (\ref{2squeezer}). Alice and Bob can then use the maximally entangled states to establish a perfect teleportation channel, allowing reliable quantum communication up to a rate $Q_2 = D_2$ \cite{bennett_mixed-state_1996}. The assistance of two-way classical communication can in general augment the quantum capacity\footnote{That does not hold true for one-way classical communication \cite{bennett_mixed-state_1996}.},  i.e., $Q_2 \geqslant Q$ \cite{bennett_mixed-state_1996}.

\begin{figure}[htb]
    \begin{center}
        \includegraphics[width=0.6\textwidth]{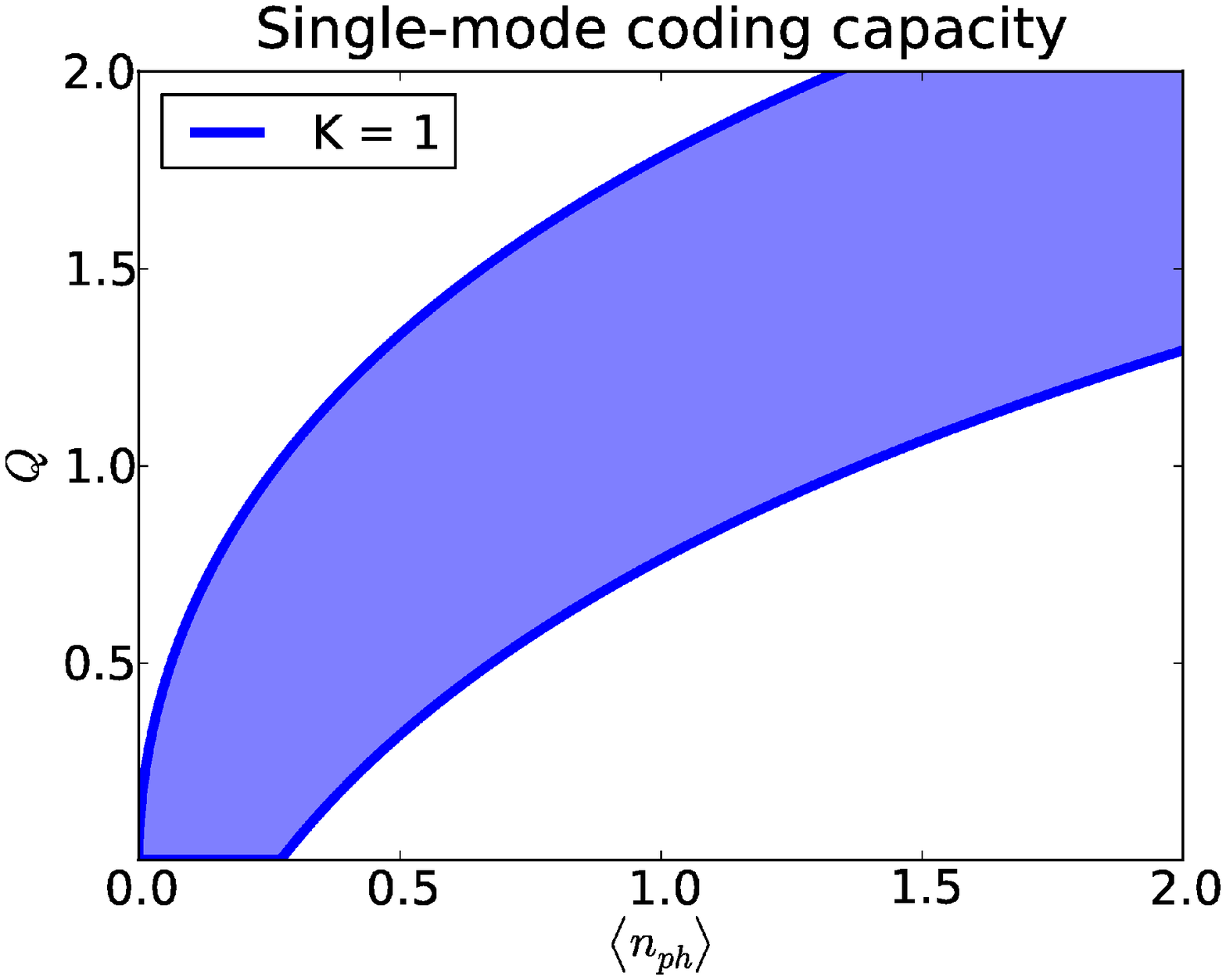}
    \end{center}
    \caption{Upper \(Q_E\) and lower \(Q_G\) bounds, in {\it q-nats}, for the quantum channel capacity of CV quantum teleportation using a single-mode EPR state when classical communication between Alice and Bob is allowed.}
    \label{fig:single_channel_capacity_classical_communication}
\end{figure}

We then compute the logarithmic negativity of the CJ state, denoted $Q_E$, which is an upper bound for $D_2$ \cite{vidal_computable_2002}. To compute the logarithmic negativity, first we have to apply the operation of partial time reversal, denoted $\Gamma$, on the CJ state (\ref{2squeezer}), which transforms the CM (\ref{2squeezerCM}) to
\begin{eqnarray}\nonumber
    \Gamma(\gamma^{BC}_s) = \frac{1}{2}
    \left( 
    \begin{array}{cccc}
    2N+\cosh{2s} & 0            & \sinh{2s} & 0         \\
    0            & 2N+\cosh{2s} & 0         & \sinh{2s} \\
    \sinh{2s}    & 0            & \cosh{2s} & 0         \\
    0            & \sinh{2s}    & 0         & \cosh{2s}
    \end{array}
    \right) \, .
\end{eqnarray}
Then we compute its symplectic eigenvalues:
\begin{equation}\fl
    d_\pm = \frac{1}{2} \sqrt{ 2N^2 + 2N\cosh{2s} + \cosh{4s} \pm ( N + \cosh{2s}) \sqrt{ 4N^2 - 2 + 2\cosh{4s} } } \, .
\end{equation}
The logarithmic negativity of the CJ state equals $\max\{ 0 , -\ln{(2d_-)} \}$. Taking the limit $s \to \infty$, after straightforward algebra, we obtain 
\begin{equation}
    Q_E = \max\{ 0 , - \ln{N} \} \, .
\end{equation}
Finally, generalizing this expression to the multi-mode setting, and putting
$N_k = e^{-2r_k}$ we obtain 
\begin{equation}
    Q_E = 2 \sum_{k=1}^n r_k \, .
    \label{eq:qe_bound}
\end{equation}
Figure \ref{fig:single_channel_capacity_classical_communication} shows the bounds $Q_G \leqslant Q_2 \leqslant Q_E$ as function of $\langle n_{ph} \rangle$. The analysis of subsections \ref{sec:multi-mode_teleportation_analysis}-\ref{sec:multi-loss} can be repeated for the quantity $Q_2$ leading to similar results: The only qualitative difference relies on the fact that the upper bound  $Q_E$ is strictly non-zero for all non vanishing values of the squeezing. In order to maximize this bound it is hence optimal to distribute the energy over as many modes as possible since there is no trade-off between the multi-mode structure and having zero quantum capacity \cite{van_enk_entangled_2005}.

\section{Optimal squeezing distributions}\label{Sec:OptDistr}
Our aim is to optimize the squeezing distribution under energy constraint. Let us denote 
\begin{equation}\label{Q1}
Q := \sum_{k=1}^K q(r_k), 
\end{equation}
($K$ integer) the function to be optimized. We want to consider general distributions, including those with an infinite number of non-zero squeezers ($K \to \infty$). To fix the ideas, we consider the case of lossless teleportation (the extension to the lossy case is straightforward). Hence, the optimization of the lower and upper bound on the lossless quantum teleportation capacity is recovered by identifying the function $q(r)$ with
\begin{equation}
q_G(r) = \max \{ 0 , 2r-1 \} \, ,
\end{equation}
or 
\begin{equation}
q_A(r) = \max \{ 0 , 2r+\log{(1-e^{-2r})} \} \, .
\end{equation}
These functions are zero if the value of $r$ is below a certain threshold. It hence follows that it is sufficient to consider a finite number of squeezers corresponding to values of the squeezing parameters above the threshold; hence we can assume without loss of generality that $K<\infty$ in (\ref{Q1}). That also allows us to substitute the function $q_G$, $q_A$ with 
\begin{eqnarray}
\tilde q_G(r) & := & 2r-1 \, , \\
\tilde q_A(r) & := & 2r+\log{(1-e^{-2r})} \, .
\end{eqnarray}
In order to optimize the quantum capacity bounds under the constraint
\begin{equation}
\langle n_{ph} \rangle = \sum_{k=1}^K \sinh^2{r_k},
\end{equation}
we introduce the Lagrange function
\begin{equation}
F( r_1, r_2, \dots r_n , \lambda) = \sum_{k=1}^K \tilde q(r_k) - \lambda \sum_{k=1}^K \sinh^2{r_k} \, ,
\end{equation}
with $\lambda$ being the Lagrange multiplier, whose value is determined by $\langle n_{ph} \rangle$, and $\tilde q$ stands for either $\tilde q_G$ or $\tilde q_A$. Differentiating with respect of $r_k$ we get the Lagrange equations
\begin{equation}
\frac{d\tilde q(r_k)}{dr_k} = \lambda \sinh{(2r_k)} \, ,
\end{equation}
which implies 
\begin{equation}
\frac{1}{\sinh{(2r_k)}} \frac{d\tilde q(r_k)}{dr_k} = \lambda \, .
\end{equation}
That means that the optimal distribution is that in which the function $\frac{1}{\sinh{(2r_k)}} \frac{d\tilde q(r_k)}{dr_k}$ is constant for all values of $k$. It hence follow that the flat distribution of the squeezing parameters is optimal. To check the uniqueness of the solution, we first note that
\begin{equation}
\frac{1}{\sinh{(2r_k)}} \frac{d\tilde q(r_k)}{dr_k} = \frac{d\tilde q(r(n_k))}{dn_k} \, ,
\end{equation}
where $r(n_k) = \mathrm{arcsinh}{\sqrt{n_k}}$. The Lagrange equations are then rewritten as follows:
\begin{equation}
\frac{d\tilde q(r(n_k))}{dn_k} = \lambda \, .
\end{equation}
A sufficient condition for the uniqueness of the solution is that the function $\tilde q(r(n_k))$ has a given concavity as function of $n_k$. The derivatives with respect to $n_k$,
\begin{eqnarray}
\frac{d\tilde q_G(r(n_k))}{dn_k} & = & \frac{1}{\sqrt{n_k(1+n_k)}} \, , \\
\frac{d\tilde q_A(r(n_k))}{dn_k} & = & \frac{e^{2\mathrm{arc}\hspace{-0.05cm}\sinh{\sqrt{n_k}}}}{e^{2\mathrm{arc}\hspace{-0.05cm}\sinh{\sqrt{n_k}}}-1} \frac{1}{\sqrt{n_k(1+n_k)}} \, ,
\end{eqnarray}
are both monotonically decreasing functions of $n_k$, which proves the concavity of $\tilde q_G(r(n_k))$, and $\tilde q_A(r(n_k))$, as functions of $n_k$.

In conclusion, we have proven that, for any given integer $K$, the flat distribution is the unique optimal squeezing distribution over the modes, as long as all individual modes feature a positive quantum channel capacity. Then, the optimal mode number $K$ can be evaluated for any given $\langle n_{ph} \rangle$, yielding the expressions presented in (\ref{eq:optimal_mode_number_flat_coding}).

\end{appendix}

\section*{References}
\bibliography{channel_capacity}

\end{document}